
\def\pd{\partial}
\def\ts{\thinspace}
\magnification=1200
\pageno=0
\parskip 3 pt plus 1pt minus 1 pt
\rightline{DTP-92/41}
\rightline{July 1992}
\vskip 2 true cm
\centerline{RELATIVISTIC RIGID PARTICLES:}
\centerline{CLASSICAL TACHYONS AND QUANTUM ANOMALIES}
\vskip 2.5 true cm
\centerline{JAN GOVAERTS%
\footnote{$\sp {\dag}$}
{\rm{Address {}from 1$\sp{\rm st}$ October 1992:}\hfill\break
\it{Institut de Physique Nucl\'eaire, Universit\'e Catholique de
Louvain,\hfill\break
B-1348 Louvain-la-Neuve (Belgium)}}}
\vskip 0.5 true cm
\centerline{\it{Department of Mathematical Sciences}}
\centerline{\it{University of Durham, Durham DH1 3LE, UK}}
\vskip 2 true cm
\centerline{Abstract}
\vskip 1 true cm
Causal rigid particles whose action includes an {\it arbitrary} dependence
on the world-line extrinsic curvature are considered. General classes of
solutions are constructed, including {\it causal tachyonic}
ones. The Hamiltonian formulation is developed in detail except for one
degenerate situation for which only partial results are given and
requiring a separate analysis. However, for otherwise generic rigid
particles, the precise specification of Hamiltonian gauge symmetries
is obtained with in particular the identification of the
Teichm$\ddot{\rm u}$ller and modular spaces for these systems. Finally,
canonical quantisation of the generic case is performed paying special
attention to the phase space restriction due to causal propagation.
A mixed Lorentz-gravitational anomaly is found in the commutator of Lorentz
boosts with world-line reparametrisations. The subspace of gauge invariant
physical states is therefore not invariant under Lorentz transformations.
Consequences for rigid strings and membranes are also discussed.
\vfill\eject
\vskip 20pt
\leftline{\bf 1. Introduction}
\vskip 15pt
Some time ago, motivated by different physical considerations,
Polyakov [1,2] proposed a modification of the ordinary Nambu-Goto
string action by including a dependence on the world-sheet
extrinsic curvature. In spite of the great deal of activity that
followed [3], a complete and exact understanding of these systems is
still lacking, especially at the quantum level. Only partial results
and educated guesses obtained through semi-classical approximation
schemes to classical solutions are available. It is generally
believed [4], though not demonstrated explicitly, that higher derivative
terms due to extrinsic curvature contributions would
render quantum unitarity impossible through physical states either of
negative norm or of energy unbounded below. Indeed, a semi-classical
analysis [5] indicates instabilities of the latter type for specific
solutions.

This situation has led some authors [6-22]
to consider the same class of actions
in the simpler case of relativistic particles. Such an investigation is
interesting not only in its own right, but it also has some relevance
to the string case in as far as particles may be viewed as collapsed
strings. However, the information available in the literature concerning
these so called rigid particles is confusing and self-contradictory.
It is thus appropriate to analyse these systems again paying greater
attention to specific issues, in particular those of canonical quantisation
not properly addressed so far.

The class of rigid particle systems considered here is described by the
general action
$$S[x\sp \mu]=\ -\mu c \int\sp {\tau_f}_{\tau_i}\ d\tau\
\sqrt{-\dot x\sp 2(\tau)}
\ F\bigl(\kappa\sp 2 K\sp 2(\tau)\bigr)\ ,\eqno(1.1)$$
where the extrinsic curvature vector is given by
$$K\sp \mu=\ {(\dot x \ddot x) \dot x\sp \mu -
\dot x\sp 2 \ddot x\sp \mu
\over (\dot x\sp 2)\sp 2}\ ,\eqno(1.2)$$
so that
$$K\sp 2=\ {\dot x\sp 2 \ddot x\sp 2 - (\dot x \ddot x)\sp 2
\over (\dot x\sp 2)\sp 3}\ .\eqno(1.3)$$
Our notations and conventions are given in Appendix A. As is also explained
there, (1.1) provides the next simplest generalisation of the ordinary
action for a relativistic scalar particle corresponding to the choice
$F(x)=1$. Indeed, (1.1) involves not only the velocities
$\dot x\sp \mu(\tau)$ but also the
accelerations $\ddot x\sp \mu(\tau)$ of the particle. By considering a
dependence
on higher $j$-torsions (see Appendix A), still higher order derivatives of
$x\sp \mu$ could be included in a consistent and systematic way.

Within the context of (1.1), two main cases have been analysed [9,16]
corresponding to the choices
$$F(x)=\ \alpha_0 \sqrt{x} + \beta_0\ ,\eqno(1.4)$$
and
$$F(x)=\ \alpha_0 x + \beta_0\ .\eqno(1.5)$$
In the present work, arbitrary choices for $F(x)$ are considered.
However, (1.4) turns out [9] to define a distinguished case, referred to as
the ``degenerate case''. In contradistinction, all other choices define the
``generic case''. Note that due to our insistence on considering (strictly)
time-like velocities only, the quantity $\kappa\sp 2 K\sp 2$ is positive
(see Appendix A). Any function $F(x)$ defined for positive arguments
is a priori allowed in (1.1). The only restriction we shall assume
here is that $F(x)$ is not constant.

First, in sect.2, the classical system is considered in some detail.
Noether (or Ward) identities [23] following from the spacetime and
world-line symmetries of (1.1) are given, and generic classes of
solutions are presented. In particular, causal but nevertheless
{\it tachyonic} solutions of constant extrinsic curvature are
found to exist always, thus generalising
the observation made in Ref.[16] for $F(x)$ given by (1.4) and (1.5). In
sect.3, we turn to the Hamiltonian formulation. The system of constraints
is analysed and the local gauge invariances associated to first-class
constraints are identified [15] -- including the associated
Teichm$\ddot{\rm u}$ller and modular spaces --
for all choices of $F(x)$ except for the degenerate case (1.4)
for which only partial results are given, leaving the complete analysis of
that case to subsequent work.
Sect.4 addresses the issue of canonical quantisation in the generic case,
{\it i.e.} for all choices of $F(x)$ different from (1.4). Due to the
restriction
on phase space following from causal propagation, first a certain change of
variables is required whereby manifest spacetime Poincar\'e covariance
-- still
a symmetry of course -- is lost. Even though the algebras of Poincar\'e and
gauge transformations are easily seen to be preserved at the quantum level,
the
quantised system turns out to have no physically acceptable interpretation,
certainly in the context of models for particle physics.
Gauge invariant physical states cannot be defined in a manner which is
at the same time consistent from the spacetime point of view. Indeed, there
appears [24] a quantum anomaly in the commutator of Lorentz boosts with
world-line
reparametrisations. Consequently, being gauge invariant becomes a frame
dependent property and in fact only the mass but not the spin of physical
states can be defined in a consistent manner. This result is derived so far
only in the generic case. In the degenerate case (1.4), the analysis of the
same issues still needs to be developed and is therefore left for future
work. However, one expects that the same conclusion concerning anomalies
would extend further to the degenerate case, and probably also to actions
including a dependence on $j$-torsions of higher order still, such as the
extrinsic torsion. Finally, in sect.5, further discussion and comments are
presented, including some consequences of our results concerning rigid
strings. Additional results secondary to the main arguments are described
in two appendices.
\vskip 20pt
\leftline{\bf 2.  Classical Solutions}
\vskip 15pt
Even though the action (1.1) is a higher order one, there is actually no
difficulty in applying the usual variational principle in order to derive
the associated Euler-Lagrange equations of motion as well as the Noether
identities and conserved quantities following from the spacetime
Poincar\'e and world-line local reparametrisation invariances of the
system. In particular, and as is typical, the equations of motion are
precisely the conservation equations for the total energy-momentum of
the system. Note also that these equations of motion are of fourth
order in $\tau$-derivatives, whose general solution thus
requires $(4D)$ integration constants. Here, we shall take for these
integration constants the initial $(\tau=\tau_i)$ and final
$(\tau=\tau_f)$ values of the coordinates $x\sp\mu(\tau)$ and
velocities $\dot x\sp \mu(\tau)$.
Of course, these integration constants will also have to obey some
constraints following from local reparametrisation invariance.

However, having in mind canonical quantisation, it would rather be more
convenient to have an action involving velocities only. Such a
redefinition of the action is readily achieved by introducing
additional degrees of freedom and
associated Lagrange multipliers whereby (1.1) is re-expressed as
$$S[x\sp \mu,q\sp \mu,\lambda\sp \mu]=\ \int\sp {\tau_f}_{\tau_i}
\ d\tau \ L(\dot x\sp \mu,q\sp \mu,\dot q\sp \mu,\lambda\sp \mu)
\ ,\eqno(2.1)$$
with
$$L=\ -\mu c\ \sqrt{-q\sp 2}\ F\bigl(\kappa\sp 2 {q\sp 2 \dot q\sp 2
- (q \dot q)\sp 2
\over (q\sp 2)\sp 3}\bigr) - \mu c\ \lambda_\mu\ (q\sp \mu - \dot x\sp \mu)
\ .\eqno(2.2)$$
Here, $q\sp \mu(\tau)$ are new degrees of freedom -- corresponding to the
velocities $\dot x\sp \mu(\tau)$ -- with a dimension of length, and
$\lambda\sp \mu(\tau)$ are dimensionless Lagrange multipliers for the
constraints $q\sp \mu=\dot x\sp \mu$.

The action (2.1) is thus our definition for the rigid particles under
consideration. As the reader is invited to verify, it should be clear that
(1.1) and (2.1) indeed describe the same classical physical system for any
given choice of $F(x)$. Corresponding to our previous remarks, the only
restrictions are that $q\sp \mu(\tau)$ is (strictly) time-like
$(q\sp 2(\tau) < 0)$
and that $F(x)$ is not constant ($F'(x)\neq 0)$.
\vskip 20pt
\leftline{\bf 2.1 Symmetries and Noether identities}
\vskip 15 pt
By construction, (2.1) possesses different symmetries. Let us consider
them in turn. Spacetime Poincar\'e transformations act as
$$x'\sp \mu=\ \Lambda\sp \mu{}_\nu\ x\sp \nu+a\sp \mu\ ,\quad
q'\sp \mu=\ \Lambda\sp \mu{}_\nu\ q\sp \nu\ ,\quad
\lambda'\sp \mu=\ \Lambda\sp \mu{}_\nu\ \lambda\sp \nu\ ,\eqno(2.3)$$
with $\Lambda\sp \mu{}_\nu$ being a Lorentz transformation and $a\sp \mu$
a constant
spacetime translation. Correspondingly, the conserved energy-momentum
$P_\mu$ and angular-momentum $M_{\mu\nu}$ are given by
$$P_\mu=\ {\pd L\over\pd\dot x\sp \mu}\ =\ \mu c\ \lambda_\mu
\ ,\eqno(2.4)$$
and
$$M_{\mu\nu}=\ L_{\mu\nu} + S_{\mu\nu}\ ,\eqno(2.5)$$
with
$$L_{\mu\nu}=\ P_\mu x_\nu - P_\nu x_\mu\ ,\quad
S_{\mu\nu}=\ {\pd L\over\pd\dot q\sp\mu} q_\nu
- {\pd L\over\pd\dot q\sp\nu} q_\mu\ .\eqno(2.6)$$
Note that the Lagrange multiplier $\lambda\sp \mu$ is essentially the
energy-momentum of the system. On the other hand, $L_{\mu\nu}$
corresponds to the covariant orbital angular-momentum, so that
$S_{\mu\nu}$ is to be interpreted as some internal spin. Such an
interpretation is indeed consistent,
as confirmed by later results. Moreover, we have
$$S\sp {\mu\nu}=\ 2\mu c \kappa \sqrt{\kappa\sp 2 K\sp 2}
\ F'(\kappa\sp 2 K\sp 2)\ K\sp {\mu\nu}_{(2)}\ ,\eqno(2.7)$$
with $K\sp {\mu\nu}_{(2)}$ given in (A.8) (and the constraint
$q\sp \mu=\dot x\sp \mu$ is to be understood of course). Thus, the
internal spin is a
direct measure of any extrinsic curvature in the world-line. In particular
for the degenerate case (1.4), the invariant $S\sp {\mu\nu}S_{\mu\nu}/2$
takes the constant value $\bigl(-(\mu c\kappa\alpha_0)\sp 2\bigr)$
irrespectively of the equations of motion. Finally, associated to the
invariance under (2.3), we have the Noether identities [23]
$${d\over d\tau}P_\mu=\ {d\over d\tau}{\pd L\over\pd\dot x\sp\mu}
\ ,\eqno(2.8)$$
$${d\over d\tau}M_{\mu\nu}=\ \sum\sp 3_{\alpha=1}
\bigl[ \bigl({d\over d\tau}{\pd L\over\pd\dot z\sp \mu_\alpha}
-{\pd L\over\pd z\sp \mu_\alpha}\bigr) z_{\alpha\nu}
- (\mu\leftrightarrow\nu)\bigr]\ ,\eqno(2.9)$$
where we set
$(z\sp \mu_1,z\sp \mu_2,z\sp \mu_3)=(x\sp \mu,q\sp \mu, \lambda\sp \mu)$
for convenience.

The action (2.1) is also invariant under reparametrisations
$(\tau \rightarrow \tilde\tau=\tilde\tau(\tau)\ts)$ which
preserve or reverse the orientation of the world-line
and leave the interval $[\tau_i,\tau_f]$ invariant (namely, $\tau_i$ and
$\tau_f$ are invariant (resp. interchanged) under orientation
preserving (resp. reversing) reparametrisations).
These transformations are defined by
$$\tilde x\sp \mu(\tilde\tau)=\ x\sp \mu(\tau)\ ,\quad
\tilde q\sp \mu(\tilde\tau)=
\ {d\tau\over d\tilde\tau}\ q\sp\mu(\tau)\ ,\quad
\tilde\lambda\sp \mu(\tilde\tau)=
\ {\rm sign}\bigl({d\tau\over d\tilde\tau}\bigr)
\lambda\sp \mu(\tau)\ .\eqno(2.10)$$
In particular, for infinitesimal
reparametrisations $\tilde\tau=\tau - \eta(\tau)$
with $\eta(\tau_i)=0=\eta(\tau_f)$, we have
$$\delta_\eta x\sp \mu=\ \eta\dot x\sp \mu\ ,\quad
\delta_\eta q\sp \mu=\ {d\over d\tau}\bigl(\eta q\sp \mu\bigr)\ ,\quad
\delta_\eta\lambda\sp \mu=\ \eta\dot\lambda\sp \mu\ .\eqno(2.11)$$
The associated generator is the canonical Hamiltonian
$$H_0=\ \sum\sp 3_{\alpha=1} \dot z\sp \mu_\alpha\
{\pd L\over\pd\dot z\sp \mu_\alpha}
- L\ ,\eqno(2.12)$$
and the corresponding Noether identities
-- Noether's second theorem [23] -- are
$$q\sp \mu\ {\pd L\over\pd\dot q\sp \mu}=\ 0\ ,\eqno(2.13)$$
$$H_0=\ q\sp \mu\ \bigl[{d\over d\tau}{\pd L\over\pd\dot q\sp \mu}
- {\pd L\over\pd q\sp \mu}\bigr]\ ,\eqno(2.14)$$
$${d\over d\tau}H_0=\ \sum\sp 3_{\alpha=1}\ \dot z\sp \mu_\alpha
\ \bigl[{d\over d\tau}{\pd L\over\pd\dot z\sp \mu_\alpha}
- {\pd L\over\pd z\sp \mu_\alpha}\bigr]\ .\eqno(2.15)$$
Note that for classical solutions, not only is $H_0$ a conserved quantity
-- Noether's first theorem (2.15) -- but it actually then also vanishes as
shown in (2.14) expressing the
invariance of solutions under local world-line reparametrisations.
On the other hand, from the relation
$${\pd L\over\pd\dot q\sp \mu}=\ -2\mu c\kappa
\ {F'(\kappa\sp 2 K\sp 2)\over\sqrt{-q\sp 2}}\ \kappa K\sp \mu
\ ,\eqno(2.16)$$
(where the relation $q\sp \mu=\dot x\sp \mu$ is again understood), it is
clear that
the identity (2.13) is equivalent to the relation $nK=0$ in (A.6) following
from the definition of the extrinsic curvature vector $K\sp \mu$ as the
variation
of the normalised tangent vector $n\sp \mu$.
\vskip 20pt
\leftline{\bf 2.2 Equations of motion}
\vskip 15pt
The equations of motion following from (2.1) are readily obtained.
Variations in $x\sp \mu$ lead to
$$\dot\lambda\sp \mu=\ 0\ ,\eqno(2.17)$$
thus expressing the conservation of the energy-momentum $P\sp \mu$.
The equation
for $x\sp \mu$ is
$$\dot x\sp \mu=\ q\sp \mu\ ,\eqno(2.18)$$
which is solved by
$$x\sp \mu(\tau)=\ x\sp \mu_i+\int\sp {\tau}_{\tau_i}
\ d\tau'\ q\sp \mu(\tau')\ .\eqno(2.19)$$
Finally, the equation for $q\sp \mu$ reduces to
$${d\over d\tau}\bigl[\kappa\sp 2\sqrt{-q\sp 2}\ F'(\kappa\sp 2 K\sp 2)
{\pd K\sp 2\over\pd\dot q\sp\mu}\bigr]=
\ \kappa\sp 2\sqrt{-q\sp 2}\ F'(\kappa\sp 2 K\sp 2)\
{\pd K\sp 2\over\pd q\sp \mu}
- {q_\mu\over\sqrt{-q\sp 2}}\ F(\kappa\sp 2 K\sp 2) +
\lambda\sp \mu\ ,\eqno(2.20)$$
with $K\sp 2$ being of course given by
$$K\sp \mu=\ {(q\dot q)q\sp \mu-q\sp 2\dot q\sp \mu\over (q\sp 2)\sp 2}\ ,
\quad
K\sp 2=\ {q\sp 2\dot q\sp 2-(q\dot q)\sp 2\over (q\sp 2)\sp 3}
\ .\eqno(2.21)$$
Clearly, these equations are solved by specifying the boundary values
$(x\sp \mu_i,q\sp \mu_i)$ and $(x\sp \mu_f,q\sp \mu_f)$ of
$(x\sp \mu(\tau),q\sp \mu(\tau)\ts)$
at $\tau=\tau_i$ and $\tau=\tau_f$ respectively. Of course, due to local
reparametrisation invariance, these boundary conditions will have
to obey a certain set of constraints. The value for $\lambda\sp \mu$ is
determined through
(2.20) and (2.19) since we must have
$$\int\sp {\tau_f}_{\tau_i}d\tau\ q\sp \mu(\tau)=\ x\sp \mu_f - x\sp \mu_i
=\ \Delta x\sp \mu\ .\eqno(2.22)$$

Before considering solutions to these equations, let us present some of the
identities that follow from them. First, given the variables
$$Q\sp \mu=\ {\pd L\over\pd \dot q_\mu}=
\ -2\mu c\kappa\ {F'(\kappa\sp 2 K\sp 2)\over\sqrt{-q\sp 2}}
\ \kappa K\sp \mu
\ ,\eqno(2.23)$$
we clearly have the identity
$$qQ=\ 0\ ,\eqno(2.24)$$
equivalent to the orthogonality condition $nK=0$ in (A.6) and the Noether
identity (2.13). On the other hand,
by projection of the equation of motion for $q\sp \mu$ on
$\lambda\sp \mu$, we also
obtain
$$qP+\ \mu c\ \sqrt{-q\sp 2}\ \bigl[
F(\kappa\sp 2 K\sp 2) - 2\kappa\sp 2 K\sp 2\
F'(\kappa\sp 2 K\sp 2)\bigr]=\ 0\ .\eqno(2.25)$$
In the degenerate case (1.4), this last relation reduces to
$$qP+\ \mu c\beta_0\ \sqrt{-q\sp 2}=\ 0\ ,\eqno(2.26)$$
whereas we then also have the further identities
$$q\sp 2 Q\sp 2+(\alpha_0\mu c\kappa)\sp 2=\ 0\ ,\quad
PQ=\ 0\ ,\quad P\sp 2 + (\mu c\beta_0)\sp 2=
\ \alpha_0\beta_0(\mu c)\sp 2\ \sqrt{\kappa\sp 2 K\sp 2}\ .\eqno(2.27)$$
These additional constraints are indicative of the distinguished
r$\hat{\rm o}$le played [9] by the degenerate case.
In particular, since $P\sp \mu$ is conserved under time evolution, the last
equality shows that in the degenerate case all classical solutions are [16]
of constant extrinsic curvature. Moreover, the same relation also
establishes that a necessary condition for the existence of classical
solutions in the degenerate case is $\beta_0\neq 0$. Indeed, $\beta_0=0$
would imply $P\sp 2=0$, but such an identity is incompatible with the
other constraints $qP=0$, $PQ=0$ and
$q\sp 2 Q\sp 2+(\alpha_0\mu c\kappa)\sp 2=0$ when only
configurations with $q\sp 2 < 0$ are considered.

In order to solve the equations of motion, it is most convenient to
consider a proper-time gauge fixing condition with
$$q\sp 2(\tau)=\ -k\sp 2\ ,\quad k\neq\ 0\ ,\eqno(2.28)$$
where $k$ is some real constant with the dimension of length.
Note that the sign of $k$ is not specified by (2.28). This ambiguity
is related [25,23] to the fact that the condition (2.28) only fixes
the gauge freedom under local reparametrisations but not under
global $Z\!\!\!Z_2$ modular transformations
corresponding to orientation reversing world-line reparametrisations.
Given the gauge (2.28), the extrinsic curvature $K\sp 2$ is simply
$$K\sp 2=\ {\dot q\sp 2\over k\sp 4}\ ,\eqno(2.29)$$
implying that the equation for $q\sp \mu$ now reads
$$\eqalign{
{2\kappa\sp 2\over |k|\sp 3}\ F'(\kappa\sp 2{\dot q\sp 2\over k\sp 4})
\ \ddot q\sp\mu +& {4\kappa\sp 4\over |k|\sp 7}\ (\dot q\ddot q)
\ F''(\kappa\sp 2{\dot q\sp 2\over k\sp 4})
\ \dot q\sp \mu=\cr
& = {1\over |k|}\ \bigl[4\kappa\sp 2{\dot q\sp 2\over k\sp 4}
\ F'(\kappa\sp 2{\dot q\sp 2\over k\sp 4})-
F(\kappa\sp 2{\dot q\sp 2\over k\sp 4})\bigr]
\ q\sp \mu + \lambda\sp \mu\ .\cr}\eqno(2.30)$$
\vskip 20pt
\leftline{\bf 2.3 Straight trajectories}
\vskip 15pt
In the proper-time gauge any straight trajectory corresponds to
$$x\sp \mu(\tau)=\ x\sp \mu_i+{\Delta x\sp \mu\over\Delta\tau}
\ (\tau - \tau_i)\ ,
\quad q\sp \mu(\tau)=\ {\Delta x\sp \mu\over\Delta\tau}\ ,\eqno(2.31)$$
with $\Delta\tau=\tau_f-\tau_i$ and
$\Delta x\sp \mu=x\sp \mu_f-x\sp \mu_i$. Thus, such
solutions may exist only if the boundary conditions are such that
$$q\sp \mu_f=\ q\sp \mu_i=\ {\Delta x\sp \mu\over\Delta\tau}\ ,\quad
(\Delta x)\sp 2 < 0\ ,\eqno(2.32)$$
with the parameter $k$ then given by
$|k|\Delta\tau=\sqrt{-(\Delta x)\sp 2}$. This is not sufficient however.
In addition, the choice for $F(x)$ must also be
such that the quantity equal to $\lambda\sp \mu$ in (2.30) is finite and
non vanishing for $q\sp \mu(\tau)$ given in (2.31), in which case this
quantity takes a value of the form
$$ A\ {\Delta x\sp \mu\over\sqrt{-(\Delta x)\sp 2}}\ ,\eqno(2.33)$$
corresponding to the value of $\lambda\sp \mu$, with $A$ thus a
dimensionless non
vanishing constant. The spacetime conserved quantities are then
$$P\sp \mu=\ {\mu c A\over\sqrt{-(\Delta x)\sp 2}}\ \Delta x\sp \mu\ ,\quad
L\sp {\mu\nu}=\ {\mu c A\over\sqrt{-(\Delta x)\sp 2}}
\ \bigl[x\sp \mu_f x\sp \nu_i - x\sp \mu_i x\sp \nu_f\bigr]\ ,\quad
S\sp {\mu\nu}=\ 0\ .\eqno(2.34)$$
This shows that straight trajectories have indeed no extrinsic curvature,
thus also no internal spin, and that the invariant mass of such solutions
is simply $\mu |A|$. Note that the sign of $A$ is related to whether
we are describing a particle as opposed to its particle (with the
particle corresponding to solutions with positive (resp. negative)
energy propagating forward (resp. backward) in time).

Therefore, provided $F(x)$ is chosen appropriatedly so that $A$ in
(2.33) is finite and non vanishing, rigid particles always have
straight trajectories as particular classical solutions. These are
precisely all classical solutions for
the ordinary scalar particle corresponding to $F(x)=1$. However, rigid
particles possess far more solutions. Nevertheless, as long as there would
exist regimes where extrinsic curvature effects are small, rigid particles
could be regarded as a viable generalisation of the ordinary scalar
particle, with even the intriguing possibility that internal spin
would follow from extrinsic curvature effects. As will become clear
later on, such a suggestion is unfortunately not tenable at the
quantum level, not even for integer spin.
\vskip 20pt
\leftline{\bf 2.4 Solutions of constant curvature}
\vskip 15pt
Obviously, it is difficult to completely solve (2.30) given an arbitrary
function $F(x)$. Nevertheless, a quite general class of solutions can be
obtained when restricting to trajectories of constant curvature $K\sp 2$.
In fact,
this class of configurations actually provides the complete solution in the
degenerate case, as was pointed out above. Note that constant values for
$K\sp 2$ are independent of the world-line parametrisation
(see Appendix A) so that we may indeed work in the proper-time gauge
(2.28) without loss of generality. Thus, given an arbitrary choice
for $F(x)$, consider configurations such that
$$K\sp 2=\ a\sp 2\ ,\quad \dot q\sp 2=\ a\sp 2 k\sp 4\ ,\eqno(2.35)$$
with $a$ being a positive constant with the dimension of
$({\rm length})\sp {-1}$. The equation (2.30) then reduces to
$$\ddot q\sp \mu=\ {k\sp 2\over 2\kappa\sp 2}\ \bigl[4 a\sp 2\kappa\sp 2 -
{F(a\sp 2\kappa\sp 2)\over F'(a\sp 2\kappa\sp 2)}\bigr]\ q\sp \mu +
{|k|\sp 3\over 2\kappa\sp 2}{1\over F'(a\sp 2\kappa\sp 2)}\ \lambda\sp\mu
\ .\eqno(2.36)$$
Three cases must therefore be considered related to the sign of the
coefficient of $q\sp \mu$ which depends on the choice for $F(x)$ and
the value of $a$. Defining
$$\alpha\sp 2=\ 4 a\sp 2\kappa\sp 2 - {F(a\sp 2\kappa\sp 2)\over
F'(a\sp 2\kappa\sp 2)}
\ ,\eqno(2.37)$$
we shall refer to these cases as being parabolic, elliptic or hyperbolic
corresponding respectively to whether $\alpha\sp 2$ is vanishing,
negative or positive. Note that $\alpha\sp 2$ is independent of $a\sp 2$
only if $F(x)=\alpha_0(x-\beta_0)\sp {1/4}$, in which case
$\alpha\sp 2=4\beta_0$.

Solving (2.36) is straightforward enough though tedious due to the
constraints (2.28) and (2.35). The general solution is constructed
as follows. Given a choice $F(x)$ and a value $a$ for the extrinsic
curvature, introduce the quantities
$$\beta=\ {\alpha\over\kappa\sqrt{2}}\ \sqrt{-q\sp 2_i}\ ,\quad
\gamma=\ {1\over 2}\ \beta \Delta\tau\ ,\eqno(2.38)$$
where $\alpha$ is a square root of $\alpha\sp 2$ in (2.37) and
$q\sp \mu_i$ is the initial boundary value for $q\sp \mu(\tau)$,
{\it i.e.} the initial velocity of the particle. Note that
$\alpha$, $\beta$ and $\gamma$ are pure imaginary in the elliptic case
and real in the hyperbolic case. Associated to this choice, a solution
of constant curvature then exists provided we have $F(x)$ and $a$
such that
$${F(a\sp 2\kappa\sp 2)\over F'(a\sp 2\kappa\sp 2)} \geq 2 a\sp 2
\kappa\sp 2\ ,\eqno(2.39)$$
and boundary conditions obeying the following constraints
$$\eqalign{q\sp 2_f=&\ q\sp 2_i < 0\ ,\qquad \Delta q\ \Delta x =\ 0\ ,\cr
(\Delta q)\sp 2=&\ a\sp 2 (\Delta\tau)\sp 2 (q\sp 2_i)\sp 2
\bigl({\cosh 2\gamma -1\over 2\gamma\sp 2}\bigr) > 0\ ,\cr
{(\Delta x)\sp 2\over (\Delta\tau)\sp 2}=&\ q\sp 2_i
-{1\over 4}(\Delta q)\sp 2{(\cosh 2\gamma +1)
\bigl({\tanh\gamma\over\gamma}\bigr)\sp 2-2
\over \cosh 2\gamma -1}\ ,\cr
q_i{\Delta x\over\Delta\tau}=&\ q\sp 2_i
-{1\over 4}(\Delta q)\sp 2{(\cosh 2\gamma +1)
\bigl({\tanh\gamma\over\gamma}\bigr)-2
\over \cosh 2\gamma -1}\ .\cr}\eqno(2.40)$$
The solution is then given as
$$\eqalignno{
x\sp \mu(\tau)=&\ x\sp \mu_i - \tilde\lambda\sp \mu (\tau - \tau_i)
+{1\over\beta}\ (q\sp \mu_i+\tilde\lambda\sp \mu)\sinh\beta(\tau-\tau_i)+\cr
&+{1\over\beta}\bigl[(q\sp \mu_f+\tilde\lambda\sp \mu)-
(q\sp \mu_i+\tilde\lambda\sp \mu)\cosh 2\gamma\bigr]
\ {\cosh\beta(\tau-\tau_i)-1\over\sinh 2\gamma}\ ,&(2.41)\cr}$$
where
$$\tilde\lambda\sp \mu=\ \bigl[\ {1\over 2}(q\sp \mu_f+q\sp \mu_i)
\bigl({\tanh\gamma\over\gamma}\bigr)-{\Delta x\sp \mu\over\Delta\tau}\ \bigr]
\bigl[\ 1-\bigl({\tanh\gamma\over\gamma}\bigr)\ \bigr]\sp{-1}=
\ {\sqrt{-q\sp 2_i}\over\mu c\alpha\sp 2 F'(a\sp 2\kappa\sp 2)}
\ P\sp \mu \ .\eqno(2.42)$$
In particular, the invariant mass of such a solution is
$$M\sp 2=\ -{1\over c\sp 2} P\sp 2=
\ \mu\sp 2\ F'\sp 2(a\sp 2\kappa\sp 2)\ \alpha\sp 2
(\alpha\sp 2-2 a\sp 2\kappa\sp 2)\ .\eqno(2.43)$$

Obviously, some comments are in order. First of all, the expressions above
are
valid only when $(\sinh 2\gamma\neq 0)$ and $(\tanh\gamma\neq\gamma)$
whenever $\gamma\neq 0$. However, a situation with $\gamma\neq 0$ and
$(\sinh 2\gamma=0)$ or $(\tanh\gamma=\gamma)$ can only occur in the
elliptic case, and the apparant singularities in the expressions
above are only a reflection of the fact that some of the integration
constants of the then periodic solutions are left
undetermined. Such a situation is analoguous to that [23]
for the ordinary harmonic oscillator when the time interval happens
to coincide with an integer multiple of the half-period. Similarly
here, no additional physical understanding is
to be gained by solving the equations whenever $(\tanh\gamma=\gamma)$ or
$(\sinh 2\gamma=0)$ with $\gamma\neq 0$. In any case, these singular
situations may always be avoided by slightly changing the value for
$\Delta\tau=\tau_f-\tau_i$.

The expressions above also define the solution when $(\alpha=0=\gamma)$
through
the appropriate limit in that variable. Correspondingly, we then have the
constraints on the integration constants
$$\eqalign{q\sp 2_f=\ q\sp 2_i &< 0\ ,\qquad \Delta q\ \Delta x=\ 0\ ,\cr
(\Delta q)\sp 2=&\ a\sp 2 (\Delta\tau)\sp 2 (q\sp 2_i)\sp 2 > 0\ ,\cr
{(\Delta x)\sp 2\over (\Delta\tau)\sp 2}=&\ q\sp 2_i-{1\over 12}
(\Delta q)\sp 2\ ,\cr
q_i{\Delta x\over\Delta\tau}=&\ q\sp 2_i-{1\over 6} (\Delta q)\sp 2
\ ,\cr}\eqno(2.44)$$
while the solution reads
$$\eqalignno{x\sp \mu(\tau)=&\ x\sp \mu_i+q\sp \mu_i(\tau - \tau_i)
+\bigl[3{\Delta x\sp \mu\over\Delta\tau}-(q\sp \mu_f+ 2 q\sp \mu_i)\bigr]
{(\tau-\tau_i)\sp 2\over\Delta\tau}+\cr
&+\bigl[(q\sp \mu_f+q\sp \mu_i)-2{\Delta x\sp \mu\over\Delta\tau}\bigr]
{(\tau-\tau_i)\sp 3\over (\Delta\tau)\sp 2}\ ,&(2.45)\cr}$$
with
$$P\sp\mu=\ {24\mu c\kappa\sp 2 F'(a\sp 2\kappa\sp 2)
\over (\Delta\tau)\sp 2 (-q\sp 2_i)\sp{3/2}}
\ \bigl[ {q\sp\mu_f+q\sp\mu_i\over 2}
- {\Delta x\sp\mu\over\Delta\tau}\bigr]\ .\eqno(2.46)$$
Obviously, the condition (2.39) is always satisfied for these parabolic
solutions, and their invariant mass vanishes identically.

Therefore, solutions of constant curvature always exist for
boundary conditions obeying (2.40) or (2.44), whatever the
choice for $F(x)$ and extrinsic curvature $a$ obeying (2.39).
In Appendix B, it is shown how
the set of solutions to (2.40) and (2.44) is indeed non empty and may
completely be specified up to arbitrary Poincar\'e transformations. In
particular, the condition (2.39) is necessary for the existence of
solutions to the constraints (2.40). For example in the degenerate
case, (2.39) requires that the parameters $\alpha_0$ and $\beta_0$ are
of the same sign, in agreement
with Ref.[16] (that reference however, does not establish the existence of
solutions to (2.40)\ts). In addition, it may be shown from (2.40) and (2.44)
that in all cases $(\Delta x)\sp 2$ is strictly negative, corresponding to a
causal observation of the particle. This property is consistent with the
conditions $\Delta x\ts\Delta q=0$ and $(\Delta q)\sp 2 > 0$. Nevertheless,
in spite of
this causality, the above solutions are all tachyonic in the
hyperbolic case! Their energy-momentum lies {\it inside} the light-cone
only in the elliptic case, and on the light-cone in the parabolic case.
For example in the degenerate case again, all solutions of curvature
$(\kappa a > \beta_0/\alpha_0 > 0)$ are tachyonic. This completes the
discussion of solutions of constant extrinsic
curvature for an arbitrary choice of $F(x)$. Solutions of non constant
curvature are more difficult to come by, unless a specific choice is
made for $F(x)$ (see for example Ref.[16] in the case (1.5)\ts).
\vskip 20pt
\leftline{\bf 3. The Hamiltonian Description}
\vskip 15pt
Corresponding to the action (2.1), the conjugate momenta are simply
$$\eqalign{
P_\mu=&\ {\pd L\over\pd\dot x\sp \mu}=\ \mu c\ \lambda_\mu\ ,\cr
Q_\mu=&\ {\pd L\over\pd\dot q\sp \mu}=\ -2\mu c\kappa
\ {F'(\kappa\sp 2 K\sp 2)\over\sqrt{-q\sp 2}}\ \kappa K_\mu\ ,\cr
\pi_\mu=&\ {\pd L\over\pd\dot\lambda\sp\mu}=\ 0\ .\cr}\eqno(3.1)$$
The Poisson bracket structure on the associated phase space is thus the
ordinary one, namely
$$\{x\sp\mu,P_\nu\}=\ \delta\sp\mu_\nu\ ,\quad
\{q\sp\mu,Q_\nu\}=\ \delta\sp\mu_\nu\ ,\quad
\{\lambda\sp\mu,\pi_\nu\}=\ \delta\sp\mu_\nu\ .\eqno(3.2)$$
Clearly, due to its local gauge invariance and the presence of Lagrange
multipliers, the Hamiltonian description of the system is subject to
constraints [23]. The following are primary constraints
$$\eqalign{
\chi\sp\mu_1=&\ \pi\sp\mu=\ 0\ ,\cr
\chi\sp\mu_2=&\ P\sp\mu-\mu c\lambda\sp\mu\ ,\cr
\phi_1=&\ q\sp\mu Q_\mu=\ 0\ .\cr}\eqno(3.3)$$
There may exist further primary constraints however. This issue is easily
settled by considering the total number of zero modes of the Hessian of the
Lagrange function (2.2). An explicit calculation shows that for a generic
function $F(x)$, the constraints (3.3) are the full set of primary
constraints. It is only in the degenerate case that an additional
primary constraint arises [9], corresponding to (compare with (2.27))
$$\chi_3=\ q\sp 2 Q\sp 2 + (\alpha_0\mu c\kappa)\sp 2 =0\ .\eqno(3.4)$$
Consequently, the Hamiltonian analysis requires a separate treatment
only for the degenerate case. All other choices of $F(x)$ may be
studied together which is done in the next section. The meaning of
the constraints above is clear. The constraints $\chi\sp\mu_1=0$ and
$\chi_2\sp\mu=0$ appear since $\lambda\sp\mu$ are Lagrange multipliers,
actually also measuring the total energy-momentum of the particle.
The constraint $\phi_1=0$ corresponds to the Noether identity (2.13)
and is thus equivalent to the relation $nK=0$ in (A.6). Therefore,
this constraint will always appear whathever the dependence on all
extrinsic $j$-torsions $(j=1,2,\cdots,D-1)$ (see Appendix A) in the
most general case. The constraint $\phi_1=0$ is not
particular to our restriction of a dependence on the 1-torsion or
extrinsic curvature only. Finally, the constraint $\chi_3=0$ is a
direct representation of the fact that for the degenerate case the
combination $x F'\sp 2(x)$ is constant (see (2.23)\ts).
\vskip 20pt
\leftline{\bf 3.1 The generic case}
\vskip 15pt
In the non degenerate case, the analysis of constraints shows that
there is only one secondary constraint which actually corresponds
to the canonical Hamiltonian. This secondary constraint is (compare
with (2.25))
$$\phi_2=\ qP + \mu c\ \sqrt{-q\sp 2}\ \Phi(q\sp 2 Q\sp 2)\ ,\eqno(3.5)$$
with the function $\Phi$ defined by
$$\Phi(q\sp 2Q\sp 2)=\ F(x_0) - 2 x_0 F'(x_0)\ ,\eqno(3.6)$$
and $x_0$ being a solution to the equation
$$x_0 F'\sp 2(x_0)=\ {-q\sp 2 Q\sp 2\over (2\mu c\kappa)\sp 2} > 0
\ .\eqno(3.7)$$
Therefore, in order to define the Hamiltonian description of generic
rigid particles, the function $F(x)$ must also be such that given
any $y > 0$ there always exists a unique $x > 0$ for which
$xF'\sp 2(x)=y$. This condition puts some restriction on the class
of acceptable functions $F(x)$ in (2.2), which is assumed to be met
in our analysis. However, one may also take the point of view that
the Hamiltonian formulation is not necessarily directly related to the
Lagrangian one in (2.2), in which case only $\Phi(q\sp 2Q\sp 2)$ needs
to be given and may be assumed to be any arbitrary {\it non constant}
function (\ts $\Phi(q\sp 2Q\sp 2)$ constant indeed corresponds to the
degenerate rigid particle). The first-class Hamiltonian -- $H_*$ in
the notation of Ref.[23] -- including only the primary first-class
constraints is obtained as
$$H_*=\ \phi_2 + u_1 \phi_1\ ,\eqno(3.8)$$
with $u_1$ being an arbitrary Lagrange multiplier for the primary
constraint $\phi_1$.

It turns out that $\chi_1\sp\mu$ and $\chi_2\sp\mu$ are second-class
constraints, while $\phi_1$ and $\phi_2$ are first-class ones. Solving for
$\chi_1\sp\mu$ and $\chi_2\sp\mu$ through the associated Dirac brackets [23]
is straightforward enough. As a result, phase space reduces to the variables
$(x\sp\mu,P_\mu;q\sp\mu,Q_\mu)$ with still the fundamental (Poisson-Dirac)
brackets in (3.3), leaving only the first-class constraints $\phi_1$ and
$\phi_2$ with the algebra
$$\{ \phi_1, \phi_2\}=\ - \phi_2\ .\eqno(3.9)$$
Hence, the total Hamiltonian generating time evolution in $\tau$ is simply
$$H_T=\ \lambda_1 \phi_1 + \lambda_2 \phi_2\ ,\eqno(3.10)$$
with $\lambda_1$ and $\lambda_2$ being arbitrary Lagrange multipliers. The
above description thus provides the Hamiltonian definition of generic rigid
particles. The associated first-order action is
$$S[x\sp\mu,P\sp\mu;q\sp\mu,Q\sp\mu;\lambda_1,\lambda_2]=
\ \int\sp{\tau_f}_{\tau_i} d\tau
\ \bigl[ \dot x\sp\mu P_\mu + \dot q\sp\mu Q_\mu
- \lambda_1\phi_1 - \lambda_2 \phi_2 \bigr]\ .\eqno(3.11)$$

The generators of spacetime Poincar\'e transformations are obviously
$P\sp\mu$ and $M\sp{\mu\nu}$ with
$$\eqalign{
M\sp{\mu\nu}=\ L\sp{\mu\nu} &+ S\sp{\mu\nu}\ ,\cr
L\sp{\mu\nu}=\ P\sp\mu x\sp\nu - P\sp\nu x\sp\mu\ ,&\quad
S\sp{\mu\nu}=\ Q\sp\mu q\sp\nu - Q\sp\nu q\sp\mu\ .\cr}\eqno(3.12)$$
Indeed, the Poisson bracket algebra of $P\sp\mu$ and $M\sp{\mu\nu}$ is
isomorphic to the Poincar\'e algebra. In addition, $L\sp{\mu\nu}$ and
$S\sp{\mu\nu}$ separately define the Lorentz algebra, $S\sp{\mu\nu}$
commutes with $P\sp\mu$ and $L\sp{\mu\nu}$, whereas $L\sp{\mu\nu}$ and
$P\sp\mu$ also define the Poincar\'e algebra. The identification of
$S\sp{\mu\nu}$ with internal spin commuting with the orbital
angular-momentum is thus consistent.

Given (3.10), it is possible to write down the Hamiltonian equations
of motion (see (3.17) below). Using the equation for $\dot q\sp\mu$
to express $Q\sp\mu$ in terms of the remaining degrees of freedom
-- this requires $\lambda_2$ to be nowhere vanishing (at least in the
interval $[\tau_i,\tau_f]$), a condition which we therefore assume
to be met throughout --, the action (3.11) reduces to
$$S[x\sp\mu,P\sp\mu;q\sp\mu;\lambda_1,\lambda_2]=\ -\mu c
\int\sp{\tau_f}_{\tau_i} d\tau
\bigl[\lambda_2 \sqrt{-q\sp 2}\ F\bigl(\kappa\sp 2
{(\dot q\sp\mu - \lambda_1 q\sp\mu)\sp 2\over \lambda\sp 2_2 (q\sp 2)\sp 2}
\bigr) + {1\over\mu c} P_\mu(\lambda_2q\sp\mu - \dot x\sp\mu)\bigr]
\ .\eqno(3.13)$$
Clearly, any dependence on $\lambda_2$ may be absorbed into a rescaling of
$q\sp\mu$. The resulting action is then still as in (3.13), with $\lambda_2$
then set equal to 1, $\lambda_1$ replaced by
$\lambda_3=\lambda_1+\dot\lambda_2/\lambda_2$ and $F(x)$ multiplied by
${\rm sign}\ts\lambda_2$. Obviously, the same remark applies to the
Hamiltonian formulation before the Lagrangian reduction of $Q\sp\mu$
is performed. Simply $q\sp\mu$ and $Q\sp\mu$ are rescaled as
$\tilde q\sp\mu=\lambda_2 q\sp\mu$ and $\tilde Q\sp\mu=Q\sp\mu/\lambda_2$
-- a transformation which preserves the canonical brackets
(3.2) --, $\lambda_2$ is set to 1, $\lambda_1$ is shifted into
$\lambda_3=\lambda_1+\dot\lambda_2/\lambda_2$ and $F(x)$ is multiplied by
${\rm sign}\ts\lambda_2$. As will become clear shortly when considering
the local Hamiltonian gauge invariances of (3.11), $\lambda_2q\sp\mu$,
$Q\sp\mu/\lambda_2$ and $\lambda_3$ are indeed combinations of degrees of
freedom invariant under transformations generated by $\phi_1$.

To conclude as to the equivalence of (3.13) with (2.1) up to the sign of
$\lambda_2$, it is still necessary to solve for $\lambda_1$ using the
constraint $\phi_1=0$, namely
$$\lambda_1=\ {q\dot q\over q\sp 2}\ .\eqno(3.14)$$
Upon rescaling of $q\sp\mu$ by $\lambda_2$, this amounts to the relation
$$\lambda_3=\ \lambda_1+{\dot\lambda_2\over\lambda_2}=
\ {\tilde q\dot{\tilde q}\over\tilde q\sp 2}\ ,\quad
\tilde q\sp\mu=\ \lambda_2 q\sp\mu\ .\eqno(3.15)$$
Note that this expression for $\lambda_3$ coincides with the value for
$\dot e/e$ with $e$ being the induced world-line einbein
$\kappa |e|=\sqrt{-\tilde q\sp 2}$. The combination
$\lambda_3=\lambda_1+\dot\lambda_2/\lambda_2$ is indeed related to an
{\it intrinsic} world-line einbein, as will soon become clear.

It is well known [23] that first-class constraints generate local
(Hamiltonian) gauge symmetries through Poisson brackets. Considering the
infinitesimal generator
$$\phi_\epsilon=\ \epsilon_1\phi_1 + \epsilon_2\phi_2\ ,\eqno(3.16)$$
with $\epsilon_1(\tau)$ and $\epsilon_2(\tau)$ arbitrary infinitesimal
functions, the associated transformations are
$$\eqalign{
\delta_\epsilon x\sp\mu=&\ \epsilon_2 q\sp\mu\ ,\qquad
\delta_\epsilon P\sp\mu=\ 0\ ,\cr
\delta_\epsilon q\sp\mu=&\ \epsilon_1 q\sp\mu + \epsilon_2
{q\sp 2 \sqrt{-q\sp 2}\over 2\mu c\kappa\sp 2 F'(x_0)}\ Q\sp\mu\ ,\cr
\delta_\epsilon Q\sp\mu=&\ -\epsilon_1 Q\sp\mu + \epsilon_2
\bigl[-P\sp\mu + \mu c\ {q\sp\mu\over\sqrt{-q\sp 2}}\ \Phi(q\sp 2 Q\sp 2)
- {q\sp 2 \sqrt{-q\sp 2}\over 2\mu c\kappa\sp 2 F'(x_0)}\ q\sp\mu\bigr]\ ,\cr
\delta_\epsilon \lambda_1=&\ \dot\epsilon_1\ ,\quad
\delta_\epsilon\lambda_2=\ \dot\epsilon_2 + \lambda_1\ \epsilon_2
-\lambda_2\epsilon_1\ .\cr}\eqno(3.17)$$
(Note that the Hamiltonian equations of motion for $(\dot x\sp\mu,
\dot P\sp\mu,\dot q\sp\mu,\dot Q\sp\mu)$ are obtained from the r.h.s.
expressions in $(\delta_\epsilon x\sp\mu,\delta_\epsilon P\sp\mu,
\delta_\epsilon q\sp\mu,\delta_\epsilon Q\sp\mu)$ through the substitutions
$\epsilon_1=\lambda_1$ and $\epsilon_2=\lambda_2$). Correspondingly, the
variation of the first-order action (3.11) is simply
$$\delta_\epsilon S=\ \int\sp{\tau_f}_{\tau_i} d\tau
\ {d\over d\tau}\bigl[\epsilon_2\bigl({\pd\phi_2\over\pd Q_\mu}Q_\mu
- \mu c \sqrt{-q\sp 2}\ \Phi(q\sp 2 Q\sp 2)\bigr)\bigr]\ .\eqno(3.18)$$
This expression is independent of $\epsilon_1$. Therefore,
when requiring the action to
be {\it exactly} invariant -- as opposed to a possible surface term -- only
$\epsilon_2(\tau)$ is restricted by the boundary conditions
$$\epsilon_2(\tau_i)=\ 0\ ,\quad \epsilon_2(\tau_f)=\ 0\ ,\eqno(3.19)$$
whereas $\epsilon_1(\tau)$ remains totally arbitrary. This is a
first indication that the (Hamiltonian) generator of world-line
reparametrisations must involve the constraint $\phi_2$.

Hence, transformations generated by $\phi_1$ alone are not related to
world-line
diffeomorphims. Actually, these transformations are easily integrated for
finite ones leading to
$$\eqalign{
x'\sp\mu(\tau)=\ x\sp\mu(\tau)\ ,&\quad P'\sp\mu(\tau)=\ P\sp\mu(\tau)\ ,\cr
q'\sp\mu(\tau)=\ \bigl[1+h(\tau)\bigr]\ q\sp\mu(\tau)\ ,&\quad
Q'\sp\mu(\tau)=\ {1\over 1+h(\tau)}\ Q\sp\mu(\tau)\ ,\cr
\lambda\sp\prime_1(\tau)=\ \lambda_1(\tau)+{\dot h(\tau)
\over 1+h(\tau)}\ ,&\quad \lambda\sp\prime_2=\ {1\over 1+h(\tau)}
\ \lambda_2(\tau)\ ,\cr
\lambda\sp\prime_3(\tau)=&\ \lambda_3(\tau)\ ,\cr}\eqno(3.20)$$
where $h(\tau)$ is an arbitrary function with the only restriction that
$(1+h(\tau)\ts)$ must be strictly positive. The local gauge symmetry
generated by $\phi_1$ -- the constraint expressing the orthogonality of
the tangent and extrinsic curvature vectors $n\sp\mu$ and $K\sp\mu$ --
thus induces a rescaling of $q\sp\mu$, $Q\sp\mu$ and $\lambda_2$ and a
shift of $\lambda_1$ such that $\tilde q\sp\mu=\lambda_2 q\sp\mu$,
$\tilde Q\sp\mu=Q\sp\mu/\lambda_2$ and
$\lambda_3=\lambda_1+\dot\lambda_2/\lambda_2$ are invariant. The existence
of this symmetry thus explains why $\lambda_2$ scales out when using
the variables $(\tilde q\sp\mu,\tilde Q\sp\mu,\lambda_3)$. Since
$\lambda_2$ is assumed to be nowhere vanishing, note that there always
exists a $h$-transformation (3.20) with $(h(\tau)=|\lambda_2(\tau)|-1)$
such that
$$\eqalign{
q'\sp\mu(\tau)=\ \sigma \tilde q\sp\mu(\tau)\ ,&\quad
Q'\sp\mu(\tau)=\ \sigma\tilde Q\sp\mu(\tau)\ ,\cr
\lambda\sp\prime_1(\tau)=\ \lambda\sp\prime_3(\tau)=
\ \lambda_3(\tau)=&\ \lambda_1(\tau)+
{\dot\lambda_2(\tau)\over\lambda_2(\tau)}\ ,\quad
\lambda\sp\prime_2(\tau)=\ \sigma\ ,\cr}\eqno(3.21)$$
where $\sigma={\rm sign}\ts(\lambda_2(\tau))=\pm 1$. Hence, without
loss of information concerning all possible gauge {\it inequivalent}
configurations of the system, we may always assume that
$\lambda_2(\tau)=\pm 1$.

To discover how transformations also involving $\phi_2$ generate the
remaining local gauge invariance of the system -- namely local world-line
reparametrisations -- it is useful to consider the variation under (3.16)
and (3.17) of the invariant combinations $\tilde q\sp\mu$,
$\tilde Q\sp\mu$ and $\lambda_3$. This suggests that infinitesimal
(Hamiltonian) world-line reparametrisations are generated by the
combination
$$\phi\sp{(R)}_\epsilon=\ \bigl[{d\over d\tau}\bigl(
{\epsilon_2\over\lambda_2}\bigr) + \lambda_1\ {\epsilon_2\over\lambda_2}
\bigr]\ \phi_1 +\epsilon_2\phi_2\ .\eqno(3.22)$$
Indeed, with the identification
$$\epsilon_2(\tau)=\ \lambda_2(\tau)\ \eta(\tau)\ ,\eqno(3.23)$$
and using the Lagrangian reduction of $Q\sp\mu$, $q\sp\mu$ and $P\sp\mu$
following from the Hamiltonian equations of motion, it is easily seen
that the transformations $\delta_\epsilon x\sp\mu$ and
$\delta_\epsilon q\sp\mu$ (and $\delta_\epsilon P\sp\mu$) induced by
$\phi\sp{(R)}_\epsilon$ in (3.22) agree with the infinitesimal
(Lagrangian) reparametrisations in (2.11). Also, note that the total
Hamiltonian (3.10) which generates evolution in $\tau$ coincides
with the generator (3.22) of reparametrisations in $\tau$ when choosing
$\epsilon_2(\tau)=\lambda_2(\tau)$, which also corresponds to
$\eta(\tau)=1$ in (3.23).

Actually, having identified the (Hamiltonian) generator of local
world-line diffeomorphisms, it becomes possible to integrate the
associated transformations to finite ones, at least for the Lagrange
multipliers $\lambda_1$ and $\lambda_2$. For this purpose, first note that
$\lambda_3$ varies under $\phi\sp{(R)}_\epsilon$ as
$$\delta_\epsilon \lambda_3=\ {d\over d\tau}\bigl[\dot\eta
+ \eta \lambda_3\bigr]\ ,\eqno(3.24)$$
where $\eta(\tau)=\epsilon_2(\tau)/\lambda_2(\tau)$. Let us introduce the
quantity
$$e(\tau)=\ e_i\ \exp\int\sp{\tau}_{\tau_i}
d\tau '\ \lambda_3(\tau ')
=\ e_i\ {|\lambda_2(\tau)|\over|\lambda_2(\tau_i)|}
\ \exp\int\sp{\tau}_{\tau_i} d\tau '\ \lambda_1(\tau ')\ ,\eqno(3.25)$$
with $e_i=e(\tau_i)$ being an arbitrary integration constant extraneous
to the system. Assuming that $e_i$ transforms under
$\phi\sp{(R)}_\epsilon$ as
$$\delta_\epsilon e_i=\ \dot\eta(\tau_i) e_i
+\eta(\tau_i)\dot e(\tau_i)\ ,\eqno(3.26)$$
(imposing the boundary condition (3.19) would imply $\eta(\tau_i)=0$\ts)
the variation of $e(\tau)$ under $\phi\sp{(R)}_\epsilon$ is simply
$$\delta_\epsilon e(\tau)=\ {d\over d\tau}\bigl[
\eta(\tau) e(\eta) \bigr]\ .\eqno(3.27)$$
(This result is of course consistent with (3.26) and the fact that
$e_i=e(\tau_i)$\ts). Hence, $e(\tau)$ is identified with an intrinsic
world-line einbein which however, couples to the system only through the
combination
$$\lambda_1 + {\dot\lambda_2\over\lambda_2}=\ \lambda_3=
\ {\dot e\over e}\ .\eqno(3.28)$$
Using this fact and the infinitesimal variations of $\lambda_1$ and
$\lambda_2$ under $\phi\sp{(R)}_\epsilon$, it is straightforward to
obtain their transformations for finite world-line reparametrisations as
$$\eqalign{
\lambda\sp\prime_1(\tau)=&\ \dot f(\tau)\lambda_1\bigl(f(\tau)\bigr)
+ {\ddot f(\tau)\over \dot f(\tau)}\ ,\cr
\lambda\sp\prime_2(\tau)=&\ \lambda_2\bigl(f(\tau)\bigr)\ ,\cr
\lambda\sp\prime_3(\tau)=&\ \dot f(\tau)\lambda_3\bigl(f(\tau)\bigr)
+{\ddot f(\tau)\over\dot f(\tau)}\ ,\cr}\eqno(3.29)$$
with $f(\tau)$ being an arbitrary function such that $f(\tau_i)=\tau_i$
and $f(\tau_f)=\tau_f$, corresponding to the world-line
diffeomorphism. Infinitesimal variations are then obtained with
$f(\tau)=\tau+\epsilon_2(\tau)/\lambda_2(\tau)$.

The algebra of constraints (3.9) being closed, the notion of
Teichm$\ddot{\rm u}$ller space is applicable. Having obtained the explicit
form for finite gauge transformations of Lagrange
multipliers, the issue of gauge fixing of the system through gauge fixing in
Teichm$\ddot{\rm u}$ller space may be addressed (for a general discussion
of these points, see Ref.[23]). Teichm$\ddot{\rm u}$ller space -- namely
[25,23] the quotient of the space of Lagrange multipliers by the group of
{\it local} (Hamiltonian) gauge transformations generated by all
first-class constraints -- is very simple for the present system. In fact,
it reduces to only two points with the simple representation
$$\lambda_1(\tau)=\ 0\ ,\qquad
\lambda_2(\tau)=\ 1 \quad {\rm or}\quad
\lambda_2(\tau)=\ -1\ .\eqno(3.30)$$
Indeed, under gauge transformations generated by $\phi_1$ and $\phi_2$,
{\it any} configuration for $(\lambda_1,\lambda_2\ts)$ -- with
$\lambda_2$ nowhere vanishing in $[\tau_i,\tau_f]$ -- is always related
to one of the two configurations (3.30). On the one hand, as was remarked
in (3.21), $\lambda_2(\tau)$ may always be set equal to
${\rm sign}\ts\lambda_2=\pm 1$ through some $h$-transformation. Note that
the configurations $\lambda_2(\tau)=\pm 1$ are invariant under world-line
diffeomorphisms in (3.29). On the other hand, any configuration of
$\lambda_3(\tau)$ may always be set to zero through some world-line
reparametrisation. As is well known, given an intrinsic einbein $e(\tau)$,
there always exists a local world-line diffeomorphism such that $e(\tau)$
is transformed into the constant configuration
$${1\over\Delta\tau}\ \int\sp{\tau_f}_{\tau_i} d\tau\ e(\tau)\ ,\eqno(3.31)$$
corresponding to the total world-line intrinsic ``length'' of the interval
$[\tau_i,\tau_f]$. Using this transformation and the correspondence (3.28),
it is clear that any $\lambda_3$ configuration is gauge equivalent to
$\lambda_3=0$. Hence finally, combining the two classes of {\it local} gauge
transformations, any configuration of $(\lambda_1(\tau),\lambda_2(\tau)\ts)$
is gauge equivalent to one of the configurations in (3.30):
Teichm$\ddot{\rm u}$ller space consists of only two points.

This conclusion also enables us to consider the issue [23]
of a complete and global,
hence admissible gauge fixing of the system through gauge fixing in
Teichm$\ddot{\rm u}$ller space. Indeed, once the configuration (3.30) is
reached, there are no further non trivial local gauge transformations
possible leaving (3.30) invariant, as is easily seen from (3.20) and
(3.29). Hence, any {\it specific} choice for
$(\lambda_1(\tau),\lambda_2(\tau)\ts)$ defines a {\it complete} gauge
fixing of the system, with the sign of $\lambda_2(\tau)$
determining which of the two Teichm$\ddot{\rm u}$ller points is selected.
Such a gauge fixing however, is not {\it global} and thus not admissible
in {\it Teichm$\ddot{\rm u}$ller space} since only {\it one} of its
two elements is singled out. Nevertheless, a gauge fixing leading to
a unique specification for $(\lambda_1(\tau),\lambda_2(\tau)\ts)$ is
global and thus admissible {\it for the system itself}. This follows
by considering the issue of modular invariance, namely transformations
under orientation {\it reversing} world-line diffeomorphisms.
As is the case [25,23] for the ordinary relativistic
particle, the Hamiltonian description of rigid particles is {\it not}
modular invariant, even though the original Lagrangian (2.1) possesses
this symmetry. Modular transformations act on phase space as
$$\eqalign{
x\sp\mu(\tau)\rightarrow x\sp\mu(\tau)\ ,&\quad
P\sp\mu(\tau)\rightarrow - P\sp\mu(\tau)\ ,\cr
q\sp\mu(\tau)\rightarrow - q\sp\mu(\tau)\ ,&\quad
Q\sp\mu(\tau)\rightarrow Q\sp\mu(\tau)\ ,\cr
\lambda_1(\tau)\rightarrow\lambda_1(\tau) ,&\quad
\lambda_2(\tau)\rightarrow - \lambda_2(\tau)\ .\cr}\eqno(3.32)$$
In particular, note that in the same way [25,23]
as for the ordinary particle, the
modular group $Z\!\!\!Z_2$ acts by exchanging the particle with its
antiparticle, since the sign of the energy $P\sp 0$ is then reversed.
Therefore, modular invariance of the original Lagrangian (2.1) is
enforced in the Hamiltonian description by requiring that
$\lambda_2(\tau)$ be, say, positive, in order to describe a rigid
particle as opposed to its antiparticle -- the latter identification
being also correlated to the sign of $F(x)$. Hence, by
specifying $(\lambda_1(\tau),\lambda_2(\tau)\ts)$ uniquely, with
$\lambda_2(\tau)$ say positive, a complete and global, hence admissible
gauge fixing of the system is effected. Such a procedure singles out
one of the two Teichm$\ddot{\rm u}$ller points, corresponding to all
Lagrange multiplier configurations which are gauge equivalent to
$$\lambda_1(\tau)=\ 0\ ,\quad \lambda_2(\tau)=\ +1\ .\eqno(3.33)$$
Note that this choice precisely corresponds to the proper-time gauge (2.28)
used in solving the equations of motion. Moreover, the configuration (3.33)
also defines modular space -- namely [25,23] the quotient of
Teichm$\ddot{\rm u}$ller space by the modular group --
which indeed reduces to a single point for the
present system due to the action (3.32) of the modular group on
Teichm$\ddot{\rm u}$ller space.
\vskip 20pt
\leftline{\bf 3.2 The degenerate case}
\vskip 15pt
The primary constraints in the degenerate case were already given in (3.3)
and (3.4). The analysis of constraints reveals two further constraints,
namely
$$\phi_2=\ qP+\mu c\beta_0\sqrt{-q\sp 2}\ ,\quad
\chi_4=\ PQ\ ,\eqno(3.34)$$
while the first-class Hamiltonian $H_*$ including only first-class primary
constraints is obtained as
$$H_*=\ \phi + u_1 \phi_1\ ,\eqno(3.35)$$
where $u_1$ is an arbitrary Lagrange multiplier for the primary constraint
$\phi_1$ and $\phi$ is defined by
$$\phi=\ \phi_2 + {1\over 2(\mu c\beta_0)}
{\bigl[P\sp 2 + (\mu c\beta_0)\sp 2\bigr]
\over (\alpha_0\mu c\kappa)\sp 2}\ \sqrt{-q\sp 2}\ \chi_3\ .\eqno(3.36)$$

As in the generic case, the constraints $\chi_1\sp\mu$ and $\chi_2\sp\mu$
turn out to be second-class ones which are easily reduced by using
Dirac brackets. Doing so leaves only the conjugate phase space degrees of
freedom $(x\sp\mu,P_\mu;q\sp\mu,Q_\mu)$ with the brackets (3.2).
The remaining constraints are then (compare with (2.26) and (2.27))
$$\eqalign{
\phi_1=&\ qQ=\ 0\ ,\quad \phi_2=\ qP+\mu c\beta_0\sqrt{-q\sp 2}=\ 0\ ,\cr
\chi_3=&\ q\sp 2 Q\sp 2 +(\alpha_0\mu c\kappa)\sp 2=\ 0\ ,\quad
\chi_4=\ PQ=\ 0\ ,\cr}\eqno(3.37)$$
with the brackets
$$\eqalign{
\{\phi_1,\phi_2\}=\ -\phi_2\ ,\quad \{&\phi_1,\chi_3\}=\ 0\ ,
\quad \{\phi_1,\chi_4\}=\ \chi_4\ ,\cr
\{\phi_2,\chi_3\}=\ 2 q\sp 2 \chi_4 + 2\mu c\beta_0 \sqrt{-q\sp 2}
\ \phi_1\ ,&\quad
\{\phi_2,\chi_4\}=\ \bigl[P\sp 2 + (\mu c\beta_0)\sp 2\bigr]
-{\mu c\beta_0\over\sqrt{-q\sp 2}}\ \phi_2\ ,\cr
\{\chi_3,\chi_4\}=&\ 2 Q\sp 2 qP\ .\cr}\eqno(3.38)$$
{}From these results, it follows that the four constraints must separate
into two first-class and two second-class constraints. Indeed, the number
of second-class constraints must be even and $\chi_4$ is clearly
second-class whereas $\phi_1$ is obviously first-class. Given (3.38),
one easily finds that the two first-class constraints are precisely
$\phi_1$ and $\phi$ given above, with the algebra
$$\{\phi_1,\phi\}=\ -\phi\ .\eqno(3.39)$$
Of course, that these are the two first-class constraints of the system
is to be expected since the first-class Hamiltonian $H_*$ in (3.35)
involves them both. The two remaining second-class constraints are then
$$\chi=\ \alpha_2\phi_2 +\alpha_3\chi_3\ ,\quad \chi_4=\ PQ\ ,\eqno(3.40)$$
where $\alpha_2$ and $\alpha_3$ are arbitrary functions on phase space such
that
$$\alpha_2\bigl[P\sp 2+(\mu c\beta_0)\sp 2\bigr]
-2(\mu c\beta_0){(\alpha_0\mu c\kappa)\sp 2\over\sqrt{-q\sp 2}}
\ \alpha_3\neq\ 0\ .\eqno(3.41)$$

The total Hamiltonian of the system is therefore
$$H=\ \lambda_1 \phi_1 + \lambda_2 \phi\ ,\eqno(3.42)$$
with $\lambda_1$ and $\lambda_2$ being arbitrary Lagrange multipliers
for the two independent first-class constraints. As in
the generic case, the constraint $\phi_1$ generates local Hamiltonian
gauge symmetries corresponding to local rescalings of $q\sp\mu$,
$Q\sp\mu$ and $\lambda_2$ and local shifts of $\lambda_1$, as given
in (3.20). On the other hand, the algebra (3.39) shows that $\phi$ is
now the generator of local world-line reparametrisations as opposed
to $\phi_2$ in the generic case. All the considerations that applied
to $\phi_2$ and its generated symmetries in the generic case should
now apply essentially to $\phi$ in the degenerate case. However, the
complete analysis of local gauge invariances associated
to the two first-class constraints $\phi_1$ and $\phi$ along the lines
of the previous section in the generic case -- including the discussion
of Teichm$\ddot{\rm u}$ller and modular spaces and gauge fixing --
is not developed here and is left for future work. The reason is that
the remaining two second-class constraints need to be included as well,
and the present author has not been able so far to do this is any
satisfactory way. On the one hand, introducing Dirac brackets associated
to $\chi$ and $\chi_4$ in (3.40) is possible but far from being elegant.
Not only are the corresponding expressions rather involved, rendering
any possible geometrical or physical insight at least very difficult
if not impossible, but the reduced Hamiltonian formulation is then no
longer manifestly Poincar\'e covariant -- a symmetry which one would
like to maintain as much as is possible. On the other, there exist other
possible approaches which in principle preserve manifest Poincar\'e
covariance by extending the formulation of the system in one or
another way -- either [26] by introduding further degrees of freedom
of opposite Grassmann parities or [27] by viewing
the second-class constraints as
resulting from the gauge fixing of some other system which encompasses the
present one and possesses first-class constraints only (these two points
of view are probably related). However, it has
not been possible so far to complete either of these two programs for
degenerate rigid particles.

For these reasons, a detailed and complete Hamiltonian analysis of
degenerate rigid particles is left for subsequent work. Such a
discussion is also necessary in order to address in a satisfactory
and complete way the canonical quantisation of these systems and in
particular their quantum physical spectrum. Therefore,
the remainder of this paper will be concerned with the canonical
quantisation of {\it generic} rigid particles only.
\vskip 20pt
\leftline{\bf 4. Canonical Quantisation of Generic Rigid Particles}
\vskip 15pt
Naively, canonical quantisation of generic rigid particles would
proceed from their Hamiltonian formulation discussed in sect.3.1.
Heisenberg commutation
relations for the fundamental degrees of freedom would simply follow
from the Poisson brackets (3.2) and in the associated representation
space of quantum states  -- necessarily  equivalent to a
wave-function representation -- physical states would be identified
as being those states annihilated by two quantum operators in direct
correspondence with the first-class constraints $\phi_1$ and $\phi_2$
for some consistent choice of normal ordering of composite operators.
However, this approach -- the one so far always adopted for rigid
particles [8-10,22,28] -- overlooks one important feature concerning
the degrees of freedom $q\sp \mu$, namely the fact that this sector of
phase space is restricted by the requirement $(-q\sp 2 > 0)$ or in other
words that $q\sp \mu$ must lie {\it inside} the light-cone. Hence, in the
same way that the canonical quantisation of the nonrelativistic particle
moving freely on the positive real axis needs some specification [29],
we must -- if we are to avoid using the more abstract methods of geometric
quantisation [30,29] -- first find a (canonical) transformation for the
restricted degrees of freedom $q\sp \mu$ and $Q\sp \mu$ such that the
new set of variables is unrestricted and preferably, is also equipped
with a canonical symplectic structure. Then, in terms of the transformed
degrees of freedom, the above quantisation program may be applied.
Such a set of transformed degrees of freedom indeed exists [24]
for rigid particles.
\vskip 20pt
\leftline{\bf 4.1 The unrestricted phase space map}
\vskip 15pt
Consider the following definitions
$$\eqalignno{y\sp 0=\ \eta\sqrt{-q\sp 2}\ ,\qquad
&R\sp 0=\ \eta\ {\bigl[-q\sp 0 Q\sp 0 +
\vec q. \vec Q \ \bigr]\over\sqrt{-q\sp 2}}\ ,&(4.1a)\cr
y\sp i=\ \eta\ {q\sp i\over\sqrt{-q\sp 2}}\ ,\qquad
&R\sp i=\ \eta\sqrt{-q\sp 2}\bigl[Q\sp i-{q\sp i\over q\sp 0}\ Q\sp 0\bigr]
\ ,&(4.1b)\cr}$$
where $\eta$ is the sign of $q\sp 0$ and $(i=1,2,\cdots,D-1)$ are space
indices. The inverse relations are
$$\eqalignno{q\sp 0=\ y\sp 0\sqrt{1+{\vec y}\ts\sp 2}\ ,\qquad
&Q\sp 0=\sqrt{1+{\vec y}\ts\sp 2}
\bigl[\ -R\sp 0+{{\vec y}.{\vec R}\over y\sp 0}\ \bigr]\ ,&(4.2a)\cr
q\sp i=\ y\sp 0\ts y\sp i\ ,\qquad
&Q\sp i=\ {R\sp i\over y\sp 0} + y\sp i
\bigl[\ -R\sp 0+{{\vec y}.{\vec R}\over y\sp 0}\ \bigr]\ . &(4.2b)\cr}$$
In geometrical terms, $y\sp 0$ measures the invariant length of the vector
$q\sp \mu$ with a sign related to whether $q\sp \mu$ lies in the forward
or in the backward light-cone, while the remaining variables $y\sp i$ are
in fact the parameters $\Lambda\sp 0{}_i$
of the Lorentz boost in the direction $\vec q\ts$
mapping the vector $q\sp \mu=(q\sp 0,\vec q\ts)$ into
the vector $(y\sp 0,\vec 0\ts)$. The variables $R\sp 0$ and
$R\sp i$ are then obtained as degrees of freedom conjugate to $y\sp 0$ and
$y\sp i$ respectively. Namely, the Poisson brackets
$$\{q\sp\mu,Q\sp\nu\}=\ \eta\sp{\mu\nu}\ ,\eqno(4.3)$$
and the following canonical brackets
$$\{y\sp 0,R\sp 0\}=1\ ,\quad \{y\sp i,R\sp j\}=\delta\sp {ij}
\ ,\eqno(4.4)$$
are mapped into one another under the transformations (4.1) and (4.2).

Clearly, the canonically conjugate degrees of freedom
$(y\sp 0,R\sp 0;y\sp i,R\sp i)$ are no longer restricted as are the
original ones $(q\sp \mu,Q_\mu)$, thereby achieving the required
properties. However, the price to pay is a loss of manifest Lorentz
covariance. Spacetime translations generated by $P\sp \mu$ and space
rotations generated by $M\sp {ij}=L\sp {ij}+S\sp {ij}$
are still manifest symmetries in the transformed representation of the
system, but this is no longer the case for Lorentz boost generators
$M\sp {0i}=L\sp {0i}+S\sp {0i}$. Indeed, while the expressions (3.12) for
$L\sp {\mu\nu}$ are not affected by the redefinitions (4.2), those for the
spin tensor become
$$S\sp {0i}=\ -R\sp i\ts\sqrt{1+{\vec y}\ts\sp 2}\ ,\quad
S\sp {ij}=\ R\sp i y\sp j - R\sp j y\sp i\ .\eqno(4.5)$$
Nevertheless, it is a straightforward calculation to check that with the
brackets (4.4), the full Poincar\'e algebra is still obtained for the
generators $P\sp \mu$ and $M\sp {\mu\nu}$ expressed in terms of the
transformed variables (4.1), thereby establishing the consistency of this
alternative Hamiltonian description of generic rigid particles (the
redefinitions (4.1) are of course also applicable in the degenerate case
but only the complete Hamiltonian analysis in that case would confirm
whether these redefinitions are also appropriate for degenerate rigid
particles). In the generic case, the first-class constraints (3.3) and
(3.5) and the associated Hamiltonian (3.10) are then given by
$$\phi_1=\ y\sp 0\ts R\sp 0\ ,\eqno(4.6)$$
and
$$\phi_2=\ y\sp 0\bigl[{\vec y}.{\vec P} - P\sp 0\sqrt{1+
{\vec y}\ts\sp 2}\bigr]
+\eta\ts\mu c\ y\sp 0 \Phi\bigl(\ts(y\sp 0 R\sp 0)\sp 2 -
({\vec y}.{\vec R}) \sp 2 -{\vec R}\sp 2\bigr)\ ,\eqno(4.7)$$
with
$$q\sp 2 Q\sp 2=\ (y\sp 0 R\sp 0)\sp 2 -
({\vec y}.{\vec R})\sp 2 -{\vec R}\sp 2\ .
\eqno(4.8)$$
{}From these expressions and the brackets (4.4), the gauge algebra (3.9)
is obviously also recovered. The present Hamiltonian formulation of
generic rigid particles is thus the one appropriate for their canonical
quantisation.
\vfill\eject
\vskip 20pt
\leftline{\bf 4.2 A mixed Lorentz-gravitational anomaly}
\vskip 15pt
Quantised generic rigid particles are thus specified by the Heisenberg
commutation relations
$$\bigl[x\sp \mu, P\sp \nu\bigr]=\ i\hbar\ \eta\sp {\mu\nu}\ ,\quad
\bigl[y\sp 0, R\sp 0\bigr]=\ i\hbar\ ,\quad \bigl[y\sp i,R\sp j\bigr]=
\ i\hbar\ \delta\sp {ij}\ ,\eqno(4.9)$$
and an abstract representation space of this algebra equipped
with an inner product for which these operators are all hermitian and
self-adjoint. Representations of this algebra are unitarily
equivalent [23] to wave-function ones either in position or in momentum
space for each pair of conjugate degrees of freedom. This determines the
space of quantum states for such systems, each of these states being
therefore of positive norm.

Turning to the ordering problem, let us first
consider the situation for the Poincar\'e generators. Clearly, $P\sp \mu$
does not require an ordering prescription. For $L\sp {\mu\nu}$ and
$S\sp {\mu\nu}$ we choose
$$L\sp {\mu\nu}=\ P\sp \mu x\sp \nu - P\sp \nu x\sp \mu\ ,\eqno(4.10)$$
and
$$S\sp {0i}=\ -{1\over 2}\bigl[\ R\sp i\sqrt{1+{\vec y}\ts\sp 2} +
\sqrt{1+{\vec y}\ts\sp 2}\ R\sp i\ \bigr]\ ,\quad S\sp {ij}=
\ R\sp i y\sp j - R\sp j y\sp i\ ,\eqno(4.11)$$
in order that these operators be hermitian and self-adjoint. Obviously,
$L\sp {\mu\nu}$ and $P\sp \mu$ generate the Poincar\'e algebra. On the
other hand, while it is clear that $S\sp {ij}$ generates the algebra of
rotations in space, it is not difficult to check that with the choice
of ordering in (4.11) the operators $S\sp {ij}$ and $S\sp {0i}$ in fact
obey the whole Lorentz algebra. Thus, the total angular-momentum
$(M\sp {\mu\nu}=L\sp {\mu\nu}+S\sp {\mu\nu})$ and energy-momentum
$P\sp \mu$ operators generate the whole Poincar\'e algebra, thereby
establishing that this algebra is anomaly free and that in spite of
the loss of manifest spacetime covariance, quantum
states of rigid particles indeed span a linear representation space for
spacetime translations and Lorentz transformations. However, since
wave-function representations of the Heisenberg algebra are single-valued,
the space of states of quantised generic rigid particles can only support
{\it integer} spin representations of the Lorentz group (though
erroneous [9,10,28],
claims for {\it half-integer} spin have indeed been made [10,14]
in the literature).

Let us now turn to the ordering problem for the first-class constraints
$\phi_1$ and $\phi_2$, a necessary prerequisite in order to define
{\it physical}, {\it i.e.} gauge invariant
states of quantum rigid particles. Again, in order to
have hermitian and self-adjoint operators, we must choose for the quantum
constraints
$$\phi_1=\ {1\over 2}\bigl[\ y\sp 0 R\sp 0 + R\sp 0 y\sp 0\ \bigr]
\ ,\eqno(4.12a)$$
and
$$\eqalignno{
\phi_2=\ y\sp 0\bigl[{\vec y}.{\vec P} - P\sp 0&\sqrt{1+
{\vec y}\ts\sp 2}\bigr]+
{1\over 2}\ \eta\ts\mu c\ \bigl[y\sp 0\Phi\bigl(
:(y\sp 0 R\sp 0)\sp 2: - :({\vec y}.{\vec R})\sp 2:
- {\vec R}\sp 2\bigr) +\cr
&+\Phi\bigl(:(y\sp 0 R\sp 0)\sp 2: - :({\vec y}.{\vec R})\sp 2:
- {\vec R}\sp 2\bigr)\ y\sp 0\ \bigr]\ ,&(4.12b)\cr}$$
where $:(y\sp 0 R\sp 0)\sp 2:$ and $:({\vec y}.{\vec R})\sp 2:$ stand for
normal ordered expressions of the corresponding operators to be specified
presently. By considering all possible orderings for the products in these
operators, one concludes that all possible choices always reduce to
expressions of the following form
$$\eqalignno{:(y\sp 0 R\sp 0)\sp 2:&=\ R\sp 0 y\sp 0 y\sp 0 R\sp 0
+ i\hbar\ A_1\ y\sp 0 R\sp 0 + \hbar\sp 2 A_2\ ,&(4.13a)\cr
:({\vec y}.{\vec R})\sp 2:&=\ R\sp i y\sp i y\sp j R\sp j
+ i\hbar\ B_1\ y\sp i R\sp i + \hbar\sp 2 B_2\ ,&(4.13b)\cr}$$
where $A_1, A_2, B_1$ and $B_2$ are undetermined free complex coefficients.
Requiring that these operators be also hermitian and self-adjoint only
leads to the restrictions
$$\eqalignno{A_1\sp *=\ - A_1\ ,&\quad A_2\sp *=\ A_2 - A_1\ ,&(4.14a)\cr
B_1\sp *=\ - B_1\ ,&\quad B_2\sp *=\ B_2 - (D-1) B_1\ .&(4.14b)\cr}$$

With these definitions, it is now possible to determine the commutation
relations for the quantum gauge algebra. One easily finds
$$\bigl[\ \phi_1,\ \phi_2\ \bigr]=\ -i\hbar\ \phi_2\ .\eqno(4.15)$$
Comparison with the classical bracket (3.9) shows that the gauge algebra is
indeed anomaly free. Therefore, both local world-line reparametrisations
and the local rescalings generated by $\phi_1$ are symmetries of quantised
generic rigid particles. {}From that point of view, it is thus
meaningful to define their quantum {\it physical} states $|\psi>$
as being the solutions to the conditions
$$\phi_1 |\psi>\ =\ 0\ ,\qquad \phi_2 |\psi>\ =\ 0\ ,\eqno(4.16)$$
thereby ensuring invariance of these states under all local gauge
symmetries including local world-line reparametrisations.

However, this
definition must also be consistent with the other symmetries of the system.
Namely, the generators of gauge symmetries {\it must commute} with those of
spacetime Poincar\'e transformations, as do the corresponding classical
brackets. Otherwise, {\it physical} states
solving (4.16) cannot define linear representations of the Poincar\'e group.
In other words, a state physical in a given reference frame would no
longer be physical in some other frame! Nor would it be possible to
define consistently the mass or the spin, or both these quantities for
{\it physical} states!

Clearly, this type of problem does not arise for the gauge
generator $\phi_1$ since
$$\eqalign{\bigl[\ L\sp {\mu\nu},\ \phi_1\bigr]=0\ ,\quad
\bigl[\ S\sp {\mu\nu},&\ \phi_1\bigr]=0\ ,\quad
\bigl[\ M\sp {\mu\nu},\ \phi_1\bigr]=0\ ,\cr
\bigl[\ P\sp \mu ,&\ \phi_1 \bigr]=0\ .\cr}\eqno(4.17)$$
Moreover, we also have for the generator of world-line reparametrisations
$$\bigl[\ P\sp \mu ,\ \phi_2 \bigr]=0\ .\eqno(4.18)$$
Therefore, at least the energy-momentum $P\sp\mu$ hence also the mass
$(M\sp 2=-P\sp 2/c\sp 2)$ of quantum
physical states are well defined observables for generic rigid particles.
To analyse the situation for the remaining commutators
$\bigl[M\sp {\mu\nu},\phi_2 \bigr]$, it is useful to decompose
$\phi_2$ in (4.12b) as $\phi_2=\chi_1+\chi_2$ with
$$\chi_1=\ y\sp 0\bigl[{\vec y}.{\vec P} - P\sp 0
\sqrt{1+{\vec y}\ts\sp 2}\bigr]\ .\eqno(4.19)$$
A simple calculation then finds that
$$\eqalignno{\bigl[\ L\sp {0i},\ \chi_1\bigr]
=&\ i\hbar (P\sp 0 y\sp 0 y\sp i - P\sp i y\sp 0\sqrt{1+{\vec y}\ts\sp 2})
=\ - \bigl[S\sp {0i},\ \chi_1\bigr]\ ,&(4.20a)\cr
\bigl[\ L\sp {ij},\ \chi_1\bigr]
=&\ i\hbar (P\sp i y\sp 0 y\sp j - P\sp j y\sp 0 y\sp i)
=\ -\bigl[\ S\sp {ij},\ \chi_1\bigr]\ ,&(4.20b)\cr}$$
leading to
$$\bigl[\ M\sp {\mu\nu},\ \chi_1\bigr]=\ 0\ ,\eqno(4.21)$$
and
$$\bigl[\ M\sp {\mu\nu},\ \phi_2\bigr]=
\bigl[\ M\sp {\mu\nu},\ \chi_2\bigr]\ .\eqno(4.22)$$
Moreover, since $L\sp {\mu\nu}$ clearly also commutes with $\chi_2$,
only the commutators of $S\sp {\mu\nu}$ with $\chi_2$ are left to be
computed. In fact, since both $y\sp 0$ and $R\sp 0$ commute with
$S\sp {\mu\nu}$, the crucial commutators to be determined are those of
$S\sp {\mu\nu}$ with
$\big(:(y\sp 0 R\sp 0)\sp 2: - :({\vec y}.{\vec R})\sp 2:
- {\vec R}\sp 2 \bigr)$.
Using the normal ordered expressions (4.13), a direct calculation shows that
$$\bigl[\ S\sp {ij},\ :(y\sp 0 R\sp 0)\sp 2: - :({\vec y}.{\vec R})\sp 2:
- {\vec R}\sp 2 \bigr]=\ 0\ ,\eqno(4.23)$$
so that finally
$$\bigl[\ M\sp {ij},\ \phi_2 \bigr]=0\ .\eqno(4.24)$$
This result is indeed to be expected owing to the manifest rotation
covariance of the quantisation procedure.

On the other hand, the commutator with Lorentz boost generators gives
$$\eqalignno{
\bigl[\ S\sp {0i},\ :(y\sp 0 R\sp 0)\sp 2: &\ - :({\vec y}.{\vec R})\sp 2:
- {\vec R}\sp 2 \bigr]=
\ {1\over 2}\ i\hbar\sp 3\ {y\sp i\over (1+{\vec y}\ts\sp 2)\sp {3/2}}\ +\cr
+\ {1\over 2}\ \hbar\sp 2\ B_1\ &\bigl[\ R\sp i{1\over\sqrt{1+
{\vec y}\ts\sp 2}}
+{1\over\sqrt{1+{\vec y}\ts\sp 2}} R\sp i\ \bigr]\ .&(4.25)\cr}$$
Hence, we certainly have for any choice of $F(\kappa\sp 2 K\sp 2)$ in the
generic case
$$\bigl[\ M\sp {0i},\phi_2\bigr] \neq 0\ .\eqno(4.26)$$
This result thus represents [24] a mixed Lorentz-gravitational anomaly in
the commutator of world-line reparametrisations and Lorentz boosts.
Generally,
this anomaly is of order $\hbar\sp 2$ unless an ordering for $\phi_2$
corresponding to $B_1=0$ in (4.13b) happens to be chosen, in which case
the anomaly is of order $\hbar\sp 3$. Therefore, given {\it any\ts}
ordering for the generator of local world-line reparametrisations,
physical states (4.16) do {\it not\ts} transform covariantly under Lorentz
boosts! The subspace of physical states (4.16) is not closed under the
action of Lorentz generators, even though these generators act covariantly
on the {\it entire} space of states and the gauge algebra is anomaly free.
\vskip 20pt
\leftline{\bf 5. Conclusions}
\vskip 15 pt
By paying closer attention to some issues  not always properly addressed
in previous works, this paper considered classical and quantum causal
rigid particles for any possible dependence of their action on the
world-line extrinsic curvature.
At the classical level, general classes of solutions of constant
extrinsic curvature were constructed, extending results of Refs.[9,16] to
any curvature dependence. These solutions always include {\it tachyonic}
ones even though the corresponding world-line trajectories always lie
{\it inside} the local light-cone in agreement with spacetime causality.
Conditions for the existence of straight trajectory solutions, {\it i.e.}
the solutions of {\it ordinary} relativistic scalar particles, were also
discussed, implying some restriction on the extrinsic curvature dependence.

The Hamiltonian formulation of these systems was also reconsidered.
Except for one degenerate situation [9] which requires a separate analysis
still to be completed -- this degenerate case represents rigid particles
whose classical trajectories are all [16] of constant curvature --
the identification [15] of local Hamiltonian gauge invariances associated
to all first-class constraints was described in
detail. In particular, Teichm$\ddot{\rm u}$ller and modular spaces for the
generic case were shown to reduce to only two and one {\it points}
respectively.
Consequently, the complete, global and thus admissible gauge fixing [23] of
the Hamiltonian description of generic rigid particles was shown to be
possible, thereby demonstrating the absence of
Gribov problems of any kind for these systems. The
degenerate case is distinguished by additional second-class constraints
rendering a manifestly Poincar\'e covariant analysis more difficult.
However, partial results in that case were presented as well.

Canonical quantisation of generic rigid particles was then considered using
the associated Hamiltonian formulation. However, due to the restriction of
causal propagation {\it inside} the light-cone, {\it i.e.} strictly
{\it time-like} velocities, a certain sector of phase space turns out to be
restricted to the {\it interior} of the light-cone. Therefore, in order to
quantise the system without having recourse to the methods of geometric
quantisation, first a certain map to an unrestricted set of
canonically conjugate phase space degrees of freedom is required. This
specific issue, which so far has never properly been addressed in the
literature,
is actually essential for a correct quantisation of rigid particles.
As a consequence of the unrestricted phase space map,
Poincar\'e covariance is no longer a manifest symmetry of the formalism.
Nevertheless, it turns out that both the spacetime  Poincar\'e algebra and
the local gauge algebra are each anomaly free even at the quantum
level. Thus, the full quantum space of states transforms covariantly under
Poincar\'e transformations and {\it physical} quantum states may be defined
as being all those states annihilated by all quantum gauge generators.
In fact, the complete space of states only supports {\it integer} spin
representations of the Lorentz group.

However, due to the necessary operator ordering of composite quantum
operators as are the gauge generators, a quantum anomaly was found [24]
for the commutator of Lorentz boosts and world-line reparametrisations.
Consequently, Lorentz boosts map {\it outside} of the {\it subspace}
of {\it physical} states even though these transformations act covariantly
on the {\it complete} space of states.
A quantum state physical in one reference frame is not necessarily
physical is some other frame! In fact, the only Poincar\'e invariant
quantum observable which is well defined for physical
states is their mass. The notion of spin has no meaning for quantum physical
states due to this mixed Lorentz-gravitational anomaly.
Therefore, quantum rigid particles cannot be considered as being consistent
models for particle physics! Their physical states cannot be defined in a
way which is compatible with the requirements of local world-line
reparametrisation
invariance and spacetime Poincar\'e covariance both at the same time. The
present quantum anomaly is quite similar to the usual mixed triangular
anomaly in four dimensions for two gravitons and one U(1) gauge boson [31],
Poincar\'e invariance playing from the world-line point of view
the r$\hat{\rm o}$le of an internal global symmetry for rigid particles.

Strictly speaking, this conclusion applies so far only for an arbitrary
dependence on the extrinsic curvature not including the degenerate
case (1.4), whose Hamiltonian description and thus canonical quantisation
still remains to be analysed. However, the same type of anomaly would
presumably be obtained in the degenerate case as well. Most probably, the
same conclusion would also extend further to theories of particles whose
actions
include a dependence on other possible $j$-torsions (see Appendix A).
If this expectation were indeed proved to be correct,
the only consistent quantum model for point-particles using as fundamental
degrees of freedom spacetime coordinates $x\sp\mu$ only -- as well as the
associated second-quantised field theories -- would simply be the
ordinary action for scalar particles, namely (1.1) with $F(x)=1$.

If these conclusions are to be any guide, one may also like to argue that
the quantum anomaly for rigid particles is the
strongest indication yet as to the probable inconsistency of quantised
rigid strings and membranes usually [4] expected on the grounds of
higher derivative couplings leading to physical states either of negative
norm or of energy unbounded below. Indeed, rigid strings and membranes
possess collapsed configurations corresponding to rigid particles.
Since quantised rigid particles are not consistent, quantised rigid strings
and membranes cannot be consistent either. Note that quantum inconsistency
of rigid particles is not related either to negative-norm
physical states nor to energy unbounded below but actually follows from
a quantum anomaly. Strictly speaking, if this type of reasoning is
justified, the conclusion applies so far only to those rigid strings and
membranes whose collapsed configurations are not degenerate rigid particles.

Specifically, consider for example the dimensional reduction [32,12]
of a rigid string in a spacetime of $(D+1)$ dimensions whose action depends
on the
world-sheet extrinsic curvature through some dimensionless function $G(x)$,
$$S[\phi\sp M]=\ -{\mu c\over\kappa}\int d\tau\ d\sigma\ \sqrt{-g}
\ G\bigl(\kappa\sp 2 \triangle\phi\sp M \triangle\phi_M\bigr)\ .\eqno(5.1)$$
Here, $\phi\sp M (M=0,1,\cdots,D)$ are the string coordinates,
$g_{\alpha\beta}=\eta_{MN}\pd_\alpha\phi\sp M\pd_\beta\phi\sp N$ is the
induced world-sheet metric ($\eta_{MN}$ is the Minkowski metric in $(D+1)$
dimensions), $\triangle$ is the Laplacian
$$\triangle\ =\ {1\over\sqrt{-g}}\ \pd_\alpha\ \sqrt{-g}
\ g\sp {\alpha\beta} \pd_\beta\ \ ,\eqno(5.2)$$
and as usual $\xi\sp {\alpha=0}=\tau$ and $\xi\sp {\alpha=1}=\sigma$
are dimensionless world-sheet coordinates
with $\alpha,\beta=0,1$. When identifying [32,12] one of the space
coordinates $\phi\sp M$ with $\sigma$ and assuming that the remaining
string coordinates $\phi\sp M \sim x\sp \mu$ are independent of $\sigma$,
(5.2) reduces to (1.1) with (the integral here is over the finite range of
$\sigma$)
$$F(x)\ =\ G(x)\ \int d\sigma\ .\eqno(5.3)$$
Thus for instance, Polyakov's rigid strings [1,12] correspond to the choice
$$F(x)=\alpha_0\ts x + \beta_0\ .\eqno(5.4)$$
Since this function does not define the degenerate case (1.4),
we must conclude from the analysis of this paper that Polyakov's rigid
strings cannot be consistent fundamental quantum theories. Of course,
this does not necessarily exclude the possible relevance of rigid string and
membrane theories -- whose actions include higher derivative couplings
characterizing the extrinsic geometry of these objects as
embedded manifolds -- as effective theories for a {\it semi-classical\ts}
approximation to the dynamics of specific solutions possessing some
extended structure in more fundamental theories.
\vskip 20 pt
\leftline{\bf Acknowledgements}
\vskip 15pt
The help of Anna Sioras in checking some of the expressions for classical
solutions is acknowledged. This work was supported through a Senior
Research Assistant position funded by the S.E.R.C.
\vfill\eject
\vskip 20pt
\leftline{\bf Appendix A}
\vskip 15pt
Consider a point particle propagating freely in a Minkowski spacetime of $D$
dimensions with metric $\eta_{\mu\nu}={\rm diag}(- + + ... +)
\ (\mu,\nu=0,1,\cdots,D-1)$. The spacetime embedding of the particle
world-line
is specified by $D$ coordinates $x\sp \mu(\tau)$
transforming as vectors under spacetime Poincar\'e symmetries and functions
of the world-line parameter $\tau$.
Correspondingly, the induced world-line metric is simply
$$\gamma(\tau)=\ -\dot x\sp 2(\tau)\ ,\eqno(A.1)$$
where as usual a dot stands for a derivative with respect to $\tau$. For
obvious physical reasons, the entire discussion is restricted to
(classical) time-like trajectories $x\sp \mu(\tau)$, namely trajectories
for which $\gamma(\tau)$ is strictly positive corresponding to strictly
time-like velocities (configurations with $\dot x\sp 2=0$ are excluded
as they correspond to a degenerate world-line metric $\gamma(\tau)$\ts).

Given the proper-time parametrisation implicitly defined by
$$ds=\ \gamma\sp {1/2}\ d\tau\ ,\eqno(A.2)$$
consider now the normalised tangent vector
$$n\sp \mu=\ {d x\sp \mu\over d s}\ =\ \gamma\sp {-1/2}\ \dot x\sp \mu
\ ,\eqno(A.3)$$
with the time-like value
$$n\sp 2(\tau)=\ -1\ ,\eqno(A.4)$$
following from the restriction $\dot x\sp 2 < 0$. Any variation in
$n\sp \mu(\tau)$ corresponds to some extrinsic curvature of the embedded
trajectory $x\sp \mu(\tau)$, namely
$$K\sp \mu =\ {d n\sp \mu\over d s}\ =\ \gamma\sp {-1/2}
\ {d \over d\tau}\bigl[\
\gamma\sp {-1/2}\ \dot x\sp \mu\ \bigr]\ .\eqno(A.5)$$
It readily follows from (A.4) that we have
$$nK=\ 0\ ,\qquad K\sp 2 \geq 0 \eqno(A.6)$$
(note that if $\dot x\sp \mu(\tau)$ were space-like rather than time-like
as assumed here, the sign of $K\sp 2$ would remain undertermined,
depending on the configuration $x\sp\mu(\tau)$\ts).

Whenever $K\sp \mu$ is non vanishing, (A.6) shows that the vectors
$n\sp \mu$ and $K\sp \mu$ are linearly independent. They thus define
a plane: the ``osculating plane''. Any variation in this plane
corresponds to some extrinsic torsion of
the embedded trajectory $x\sp \mu(\tau)$. Clearly, this extrinsic
torsion is non
vanishing whenever the vectors $(n\sp \mu,K\sp \mu)$ and
$(d n\sp \mu/ d s=K\sp \mu,d K\sp \mu/ d s)$ are linearly independent.
An orthogonal
decomposition of $d K\sp \mu/ d s$ with respect to $(n\sp \mu, K\sp \mu)$
then leads to
the following definition of extrinsic torsion ($\kappa$ is some arbitrary
physical positive constant with a dimension of length)
$$T\sp \mu=\ {d K\sp \mu\over d s} - ({d\over d s}\ln
\sqrt{\kappa\sp 2 K\sp 2})\ K\sp \mu
- K\sp 2 n\sp \mu\ .\eqno(A.7)$$
Equivalently, consider the normalised two-form characterizing the linear
independence of $n\sp \mu$ and $K\sp \mu$ when $K\sp \mu\neq 0$
$$K\sp {\mu\nu}_{(2)}=\ n\sp \mu\ {K\sp \nu\over\sqrt{K\sp 2}}
- n\sp \nu\ {K\sp \mu\over\sqrt{K\sp 2}}\ \ .\eqno(A.8)$$
Any variation in $K\sp {\mu\nu}$ corresponds to some extrinsic torsion.
Indeed,
one has
$$T\sp {\mu\nu}_{(2)}=\ {d K\sp {\mu\nu}_{(2)}\over d s}\ =
\ {1\over\sqrt{K\sp 2}}
\ \bigl[\ n\sp \mu T\sp \nu - n\sp \nu T\sp \mu\ \bigr]\ ,\eqno(A.9)$$
with $T\sp \mu$ given in (A.7). {}From these definitions, it follows
again that
$$nT =\ 0\ ,\quad KT =\ 0\ ,\quad T\sp 2 \geq 0\ .\eqno(A.10)$$

Clearly, the type of considerations above generalises easily, leading
to the definition of higher order quantities further characterizing
the extrinsic geometry of the embedded world-line $x\sp \mu(\tau)$.
For example, when $K\sp \mu$ and $T\sp \mu$ are non vanishing, (A.6)
and (A.10) establish the linear independence of the vectors
$(n\sp \mu,K\sp \mu,T\sp \mu)$ which thus span an
``osculating 3-volume''. Any variation in this 3-volume represents some
further extrinsic ``3-torsion'' of the embedded world-line. Obviously,
in a spacetime of $D$ dimensions, there only exist $D$ such vector
quantities, each corresponding to some extrinsic ``$j$-torsion''
$(j=0,1,\cdots,D-1)$ characterizing the extrinsic geometry of the
spacetime trajectory $x\sp \mu(\tau)$, with the 0-torsion, 1-torsion
and 2-torsion being the tangent, curvature and torsion vectors
$n\sp \mu$, $K\sp \mu$ and $T\sp \mu$ respectively. In terms of the
coordinates $x\sp \mu(\tau)$, the extrinsic $j$-torsion
$T\sp \mu_{(j)}$ involves derivatives with respect to $\tau$ of
order $(j+1)$ and less.

By construction, all $j$-torsions $T\sp \mu_{(j)}$ are manifestly covariant
vectors for spacetime Poincar\'e transformations, and invariant quantities
under local, {\it i.e.} orientation preserving world-line
reparametrisations. For global, {\it i.e.} orientation reversing
world-line reparametrisations, all $(2j)$-torsions
($j=0,1,\cdots,[(D+1)/2]-1$; $[x]$ denotes the integer part of
$x$) change sign whereas all
$(2j+1)$-torsions $(j=0,1,\cdots,[D/2]-1)$ are invariant. However, since
by definition all $j$-torsions are mutually orthogonal, the only possible
independent Poincar\'e invariants are simply the quantities
$$T\sp 2_{(j)}=\ \eta_{\mu\nu}\ T\sp \mu_{(j)}\ T\sp \nu_{(j)}\ ,
\quad j=0,1,\cdots,D-1\ ,\eqno(A.11)$$
with
$$T\sp 2_{(0)}=\ n\sp 2\ =\ -1\ ,\quad T\sp 2_{(j)} \geq 0\ ,
\quad j=1,2,\cdots,D-1\ .\eqno(A.12)$$
Clearly, all $T\sp 2_{(j)}$ are also invariant under local {\it and} global
world-line reparametrisations.

Hence, the most general manifestly Poincar\'e and world-line
reparametrisation
invariant action that can be constructed using only the world-line scalar
``fields'' $x\sp \mu(\tau)$ is of the form $(\tau_i < \tau_f)$
$$S[x\sp \mu]=\ -\mu c\int\sp {\tau_f}_{\tau_i} d\tau
\ \sqrt{-\dot x\sp 2}\ {\cal F}(\kappa\sp {2j} T\sp 2_{(j)})\eqno(A.13)$$
(actually for $D=2$, a 2-cocycle term proportional to
$\epsilon_{\mu\nu} x\sp \mu\dot x\sp \nu$ could be added [33] to (A.13).
Such a term
however, breaks spacetime parity and changes sign under global world-line
reparametrisations). It is understood that the coordinates
$x\sp \mu(\tau)$ have a
dimension of length, with in particular $x\sp 0(\tau)=c t(\tau)$
-- $c$ being the speed of light and $t(\tau)$ the physical time.
Thus, $\mu$ and $\kappa$ in
(A.13) are fundamental positive physical constants characteristic of the
system, with a dimension of mass and length respectively. Finally,
${\cal F}$ is some specific dimensionless function of the $j$-torsion
invariants $T\sp 2_{(j)} (j=1,2,\cdots,D-1)$. Clearly, $\mu$ sets the
mass scale of the sytem and $\kappa$ the intrinsic length scale in the
world-line. For example, the ordinary case of the relativistic scalar
particle of mass $m$ corresponds to
the choices ${\cal F}=\pm 1$ and $\mu=m$ -- with the sign of ${\cal F}$
distinguishing between the particle and its antiparticle.
\vskip 20pt
\leftline{\bf Appendix B}
\vskip 15pt
In this appendix, we show that there do indeed exist solutions to the
constraints (2.40) and (2.44) for the boundary conditions
$(x\sp\mu_f,x\sp \mu_i,q\sp \mu_f,q\sp \mu_i)$. Taking advantage of
Poincar\'e invariance
at the classical level, any solution to (2.40) is equivalent, under a
spacetime translation and a Lorentz transformation, to the following
choice of boundary conditions
$$\eqalign{x\sp \mu_i=\ (0,0,0,\vec 0)\ ,
&\quad {\Delta x\sp \mu\over\Delta\tau}=\ (\delta,0,0,\vec 0)\ ,\cr
q\sp \mu_i=\ (q\sp 0_i,-{1\over 2}\Delta q\sp 1,q\sp {\mu=2}_i,\vec 0)\ ,
&\quad q\sp \mu_f=\ (q\sp 0_i,{1\over 2}\Delta q\sp 1,
q\sp {\mu=2}_i,\vec 0)\ ,\cr}\eqno(B.1)$$
where
$$\eqalign{
\delta&=\ \epsilon_1 |k| \bigl[\ 1+({a k\Delta\tau \over 2})\sp 2
\ {(\cosh 2\gamma + 1)\bigl({\tanh \gamma\over\gamma}\bigr)\sp 2 - 2
\over 2\gamma\sp 2}\ \bigr]\sp {1/2}\ ,\cr
q\sp 0_i&=\ {k\sp 2\over\delta}
\bigl[\ 1+({a k\Delta\tau \over 2})\sp 2\ {(\cosh 2\gamma+1)
\bigl({\tanh\gamma\over\gamma}\bigr)-2\over 2\gamma\sp 2}\ \bigr]\ ,\cr
{1\over 2}\Delta q\sp 1&=\ \epsilon_2 k
\ ({a k \Delta\tau \over 2})\sqrt{{\cosh 2\gamma-1\over 2\gamma\sp 2}}\ ,\cr
q\sp {\mu=2}_i&=\ \epsilon_1 \epsilon_3\ {k\sp 2\over\delta}
\ {1\over a\kappa\sqrt{2}}\ \bigl({a k\Delta\tau\over 2}\bigr)\sp 2
\ \bigl[ {(\cosh 2\gamma+1)\bigl({\tanh\gamma\over\gamma}-1\bigr)\sp 2
\over 2\gamma\sp 4}
\bigl({F(a\sp 2 \kappa\sp 2)\over F'(a\sp 2 \kappa\sp 2)}
- 2 a\sp 2\kappa\sp 2\bigr) \bigr]\sp {1/2}\ .\cr}\eqno(B.2)$$
Here, $\epsilon_1=\pm 1$, $\epsilon_2=\pm 1$, $\epsilon_3=\pm 1$ are
arbitrary sign factors -- corresponding to arbitrary spacetime
reflections -- and $k$ is such that
$\eta_{\mu\nu} q\sp \mu_i q\sp \nu_i=-k\sp 2$ (see (2.28)).
The remainder of the notation is defined in sect.2.4.

In the limit $\gamma=0$ corresponding to the parabolic case, the
expressions in (B.2) reduce to
$$\eqalign{
\delta&=\ \epsilon_1\ |k|\bigl[1+{1\over 12}\
(a k\Delta\tau)\sp 2\bigr]\sp {1/2}\ ,\cr
q\sp 0_i&=\ {k\sp 2\over\delta}
\ \bigl[\ 1+{1\over 6}\ (a k\Delta\tau)\sp 2\ \bigr]\ ,\cr
{1\over 2}\Delta q\sp 1&=\ \epsilon_2 k
\bigl({a k\Delta\tau\over 2}\bigr)\ ,\cr
q\sp {\mu=2}_i&=\ \epsilon_1 \epsilon_3\ {k\sp 2\over\delta}
\ {1\over 12}(a k\Delta\tau)\sp 2\ ,\cr}\eqno(B.3)$$
or equivalently
$$q\sp 0_i=\ \epsilon_1 |\delta| (1+\eta)\ ,\quad
{1\over 2}\Delta q\sp 1=\ \epsilon_2 |\delta|\ \sqrt{3\eta}\ ,\quad
q\sp {\mu=2}_i=\ \epsilon_3 |\delta| \eta\ ,\eqno(B.4)$$
where
$$\eta=\ {{1\over 12}(a k\Delta\tau)\sp 2\over
1+{1\over 12}(a k\Delta\tau)\sp 2}\ ,\quad 0 < \eta < 1\ .\eqno(B.5)$$
In this latter parametrisation, we then have
$$|k|=\ |\delta |\ \sqrt{1-\eta}\ ,\quad a=\ {2\ \sqrt{3\eta}
\over \Delta\tau\ |\delta |\ (1-\eta)}\ \ .\eqno(B.6)$$

Therefore, up to arbitrary Poincar\'e transformations, any solution
to (2.40) and (2.44) is parametrised by the choice of the value $a$ for
the extrinsic curvature and the value for
$\eta_{\mu\nu}q\sp\mu_i q\sp \nu_i=-k\sp 2$. In other
words, given any initial time-like velocity $q\sp \mu_i$ and an extrinsic
curvature value $a$, a solution of constant extrinsic curvature may always
be found for any choice of function $F(x)$ such that (2.39) is obeyed.
\vfil\eject
\vskip 20pt
\leftline{\bf REFERENCES}
\vskip 20pt
\frenchspacing
\item{[1]} A. M. Polyakov, Nucl. Phys. {\bf B268} (1986) 406.
\item{[2]} H. Kleinert, Phys. Lett. {\bf B174} (1986) 335.
\item{[3]} For a recent review, see \hfil\break
G. German, Mod. Phys. Lett. {\bf A6} (1991) 1815.
\item{[4]} J. Polchinski and Z. Yang, {\sl ``High Temperature Partition
Function of the Rigid String''}, Texas/Rochester preprint UTTG-08-92,
UR-1254, ER-40685-706.
\item{[5]} E. Braaten and C. K. Zachos, Phys. Rev. {\bf D34} (1987) 1512.
\item{[6]} R. D. Pisarski, Phys. Rev. {\bf D34} (1986) 670.
\item{[7]} C. Battle, J. Gomis and N. Roman-Roy,
J. Phys. {\bf A21} (1988) 2693.
\item{[8]} V. V. Nesterenko, J. Phys. {\bf A22} (1989) 1673;
Theor. Math. Phys. {\bf 86} (1991) 168; Mod. Phys. Lett. {\bf A6} (1991) 719.
\item{[9]} M. S. Plyushchay, Mod. Phys. Lett. {\bf A3} (1988) 1299;
Int. J. Mod. Phys. {\bf A4} (1989) 3851;
Phys. Lett. {\bf B253} (1991) 50.
\item{[10]} M. S. Plyushchay, Mod. Phys. Lett. {\bf A4} (1989) 837;
{\it ibid} {\bf A4} (1989) 2747;
Phys. Lett. {\bf B235} (1990) 47; {\it ibid} {\bf B236} (1990) 291;
{\it ibid} {\bf B243} (1990) 383; {\it ibid} {\bf B248} (1990) 107;
{\it ibid} {\bf B248} (1990) 299; {\it ibid} {\bf B262} (1991) 71;
{\it ibid} {\bf B280} (1992) 232; Nucl. Phys. {\bf B362} (1991) 54.
\item{[11]} A. Dhar, Phys. Lett. {\bf B214} (1988) 75.
\item{[12]} J. Grundberg, J. Isberg, U. Lindstr\"om and H. Nordstr\"om,
Phys. Lett. {\bf B231} (1989) 61.
\item{[13]} J. P. Gauntlett, K. Itoh and P. K. Townsend,
Phys. Lett. {\bf B238} (1990) 65; \hfil\break
J. P. Gauntlett and C. F. Yastremiz,
Class. Quantum Grav. {\bf 7} (1990) 2089; \hfil\break
J. P. Gauntlett, Phys. Lett. {\bf B272} (1991) 25.
\item{[14]} M. Pav\v si\v c, Phys. Lett. {\bf B205} (1988) 231;
{\it ibid} {\bf B221} (1989) 264.
\item{[15]} M. Huq, P. I. Obiakor and S. Singh,
Int. J. Mod. Phys. {\bf A5} (1990) 4301.
\item{[16]} H. Arodz, A. Sitarz and P. Wegrzyn,
Acta Phys. Polonica {\bf B20} (1989) 921.
\item{[17]} T. Dereli, D. H. Harley, M. \"Onder and R. W. Tucker,
Phys. Lett. {\bf B252} (1990) 601.
\item{[18]} D. Zoller, Phys. Rev. Lett. {\bf 65} (1990) 2236.
\item{[19]} G. Fiorentini, M. Gasperini and G. Scapetta,
Mod. Phys. Lett. {\bf A6} (1991) 2033.
\item{[20]} A. M. Polyakov, Mod. Phys. Lett. {\bf A3} (1988) 325.
\item{[21]} S. Iso, C. Itoi and H. Mukaida,
Phys. Lett. {\bf B236} (1990) 287; Nucl. Phys. {\bf B346} (1990) 293.
\item{[22]} Yu. A. Kuznetsov and M. S. Plyushchay, {\sl ``The Model of
the Relativistic Particle with Curvature and Torsion''},
Protvino preprint IHEP 91-162 (October 1991), and reference therein.
\item{[23]}For a recent review, see \hfil\break
J. Govaerts, {\it Hamiltonian Quantisation and Constrained Dynamics},
Lecture Notes in Mathematical and Theoretical Physics {\bf 4}
(Leuven University Press, Leuven, 1991).
\item{[24]} J. Govaerts, {\sl ``A Quantum Anomaly for Rigid Particles''},
Durham preprint DTP-92/37 (July 1992), hepth/9207068.
\item{[25]} J. Govaerts, Int. J. Mod. Phys. {\bf A4} (1991) 173;
{\it ibid} {\bf A4} (1991) 4487.
\item{[26]} I. A. Batalin and E. S. Fradkin, Phys. Lett. {\bf B180}
(1986) 157; Nucl. Phys. {\bf B279} (1987) 514; \hfil\break
I. A. Batalin, E. S. Fradkin and T. E. Fradkina,
Nucl. Phys. {\bf B314} (1989) 158.
\item{[27]} K. Harada and H. Mukaida, Z. Phys. {\it C48} (1990) 151.
\item{[28]} J. Isberg, U. Lindstr\"om and H. Nordstr\"om,
Mod. Phys. Lett. {\bf A5} (1990) 2491.
\item{[29]}C. J. Isham, in {\it Relativity, Groups and Topology II},
Les Houches 1983, eds. B. S. DeWitt and R. Stora (North Holland,
Amsterdam, 1984), p. 1162.
\item{[30]} See for example \hfil\break
N. M. J. Woodhouse, {\it Geometric Quantisation}
(Oxford University Press, Oxford, 1980).
\item{[31]} L. Alvarez-Gaum\'e and E. Witten,
Nucl. Phys. {\bf B234} (1983) 269.
\item{[32]} M. J. Duff, P. S. Howe, T. Inami and K. S. Stelle,
Phys. Lett. {\bf B191} (1987) 70.
\item{[33]} D. R. Grigore, {\sl ``A Derivation of the Nambu-Goto Action
from Invariance Principles''}, preprint CERN-TH.6101/91 (May 1991).
\end